\documentclass[nofootinbib,aps,prc,twocolumn,superscriptaddress]{revtex4-2}
\usepackage{graphicx}
\usepackage{dcolumn}
\usepackage{bm}
\usepackage{bbm}
\usepackage{color} 
\RequirePackage{booktabs}
\usepackage{longtable}
\usepackage[caption=false]{subfig}

\usepackage{amsfonts}
\usepackage{amssymb}
\usepackage{amsmath}
\usepackage{amsxtra}

\makeatletter
\newcommand{\colorcaption}[2][]{%
  \begingroup%
  \renewcommand{\@caption@fignum@sep}{ (color online). }%
  \caption[#1]{#2}%
  \endgroup%
}
\makeatother

\begin{document}

\title{Converging Many-Body Perturbation Theory for \textit{Ab Initio} Nuclear Structure: \\
II. Brillouin-Wigner Perturbation Series for Open-Shell Nuclei}

\date{\today}

\author{Zhen~Li} 
\email[]{lizhen@cenbg.in2p3.fr; physzhenli@outlook.com}
\affiliation{LP2IB (CNRS/IN2P3 -- Universit\'e de Bordeaux), 33170 Gradignan, France }
\author{Nadezda~A.~Smirnova} 
\email[]{nadezda.smirnova@u-bordeaux.fr}
\affiliation{LP2IB (CNRS/IN2P3 -- Universit\'e de Bordeaux), 33170 Gradignan, France }

\begin{abstract}
Brillouin-Wigner (BW) perturbation theory is developed for both ground and excited states of open-shell nuclei. We show that with optimal partitioning of the many-body Hamiltonian proposed earlier by the authors [Z. Li and N. Smirnova, arXiv:2306.13629], one can redefine the BW perturbation series for a given state of the effective Hamiltonian in a small $\mathbbm{P}$-space to be converging under the condition that the energy of this state is below the lowest eigenvalue of the Hamiltonian matrix block belonging to the complement of the $\mathbbm{P}$-space, characterized by the same good quantum numbers as the state under consideration. Specifically, the BW perturbative calculations for the lowest $J^\pi$ states are always converging due to the variational principle. This property does hold for both soft and hard internucleon interactions in the harmonic oscillator basis. To illustrate this method and check the convergence behavior, we present numerical studies of low-energy spectra of $^{5,6,7}$Li using the Daejeon16 and bare N$^3$LO potentials.
\end{abstract}

\bibliographystyle{apsrev4-2}

\maketitle

\section{Introduction}

Solving the many-body Schr\"odinger equation for atomic nuclei is the key to the \textit{ab initio} study of nuclear structure. The last two decades have witnessed huge progress in the developments of nuclear \textit{ab initio} many-body methods~\cite{Hergert2020_review}, accompanied by the success of chiral effective field theory (EFT) potentials~\cite{Epelbaum2009,Machleidt2011} and renormalization techniques~\cite{Bogner2010SRG}. For few-nucleon systems and very light nuclei, Faddeev equations~\cite{Faddeev1961,Yakubowsky1967}, hyperspherical harmonics method~\cite{Ballot1980,Barnea1999} and quantum Monte Carlo methods~\cite{Pieper2001} are able to provide very accurate calculations. The full configuration interaction method, no-core shell model (NCSM)~\cite{Barrett2013}, gives very detailed information on nuclear states and transitions at low energies, by diagonalizing the Hamiltonian matrix calculated in a sufficiently large model space spanned by many-body states. However, with the current computing power NCSM can only reach nuclei to oxygen or so at the moment, due to the rapid growth of the model-space dimension as the nucleon number increases. For nuclei heavier than oxygen, specific approximations need to be applied in the calculation, which has led to a variety of many-body methods, such as many-body perturbation theory (MBPT)~\cite{Coraggio2009,Tichai2020}, coupled cluster method~\cite{Hagen2014}, in-medium similarity renormalization group~\cite{Hergert2016_IMSRG}, self-consistent Green's function~\cite{Dickhoff2004}, etc. These methods have achieved a lot of success in the \textit{ab initio} study of intermediate-mass nuclei~\cite{Hergert2020_review} and even heavy nuclei like lead~\cite{Hu2022ab}. 

Initiated from the 1950s~\cite{Brueckner1955c,Brueckner1955a,Bethe1956,Goldstone1957}, MBPT has been extensively applied to the calculation of the nuclear ground-state energy~\cite{Tichai2020} and to the construction of the effective Hamiltonian used in the valence-space nuclear shell model calculation~\cite{KuoOsnes1990,MHJ1995,Coraggio2009,Coraggio2023_PPNP}. Furthermore, MBPT has also become a precious tool for the solution of atomic and molecular problems~\cite{Kelly1963,Kelly1964a,Kelly1964b,Bartlett1974a,Bartlett1974b,Bartlett1981,Shavitt}. In spite of extensive applications, MBPT is brought into doubt by the issue of order-by-order convergence in powers of interactions, especially for atomic nuclear systems, where (strong) internucleon interactions are used. Previously most attention has been attracted to the Rayleigh-Schr\"{o}dinger (RS) formulation of MBPT, because of its energy-independence feature and a convenient implementation via a diagrammatic approach~\cite{Goldstone1957,Shavitt,Kuo1971,Ellis1977}. However, a huge number of diagrams required at higher orders limited the application of this diagrammatic approach up to usually 3rd order in nuclear structure calculations~\cite{Coraggio2003,Coraggio2005,Roth2006,Gunther2010,Hu2016,Tichai2020,MHJ1995,Coraggio2009}, which makes it impossible for the study of order-by-order convergence within this diagrammatic framework. The study of the convergence properties became possible with the implementation of the algebraic recursive calculation of the RS perturbation series terms~\cite{Lowdin1965,Laidig1985}. Extensive calculations for molecular systems with various partitioning schemes reported cases of converging or diverging series~\cite{Perrine2005}. Similar calculations have also been performed for atomic nuclei~\cite{Roth2010,Langhammer2012_Pade,Tichai2016}, to which we focus our attention in the present paper. Although the recursive method was still restricted to relatively light nuclei, some convergence properties were addressed in those studies after numerical calculations. With the use of softened chiral EFT potentials, the authors~\cite{Roth2010,Langhammer2012_Pade,Tichai2016} concluded that the perturbative calculation of nuclear ground and excited states typically diverges in the HO basis, and only converges in the Hartree-Fock (HF) basis. However, harder potentials can still spoil the convergence in the HF basis~\cite{Tichai2016}. 

In our recent work~\cite{BWMBPT2023}, we chose to explore the energy-dependent BW perturbation series for ground states of closed-shell nuclei. We showed that the with an optimal Hamiltonian partitioning, the convergence criterion of a perturbative expansion for the ground-state energy can always be satisfied. This property holds due to the variational principle and does not depend on the choice of the (HO or HF) basis or the choice of the (hard or soft) internucleon interaction. In the present work, we generalize these ideas to ground and excited states of open-shell nuclei and investigate the corresponding convergence behavior. This paper is organized as follows. In Sec.~\ref{sec:formalism} we present the formalism of BW perturbation theory, the general partitioning of Hamiltonian, the convergence criterion, and the $\hat{K}$-box vertex to reach high order terms of the perturbation series. In Sec.~\ref{sec:results} we show the results of calculations for $^{5,6,7}$Li. The last section summarizes the results and formulates perspectives.

\section{Formalism: Hamiltonian partitioning and convergence criterion}\label{sec:formalism}

In this section, we generalize the BW formulation of MBPT given in Ref.~\cite{BWMBPT2023} to open-shell nuclei.  

\subsection{$\mathbbm{P}$-Space Eigenvalue Problem}\label{sec:p_space_eigenvalue_problem}

Ideally, one searches to solve the eigenvalue problem for an intrinsic Hamiltonian $H$ in a large, but finite-dimensional (denoted as $d$-dimensional) model space
\begin{eqnarray}\label{eq:full_space_schrodinger_equation}
H |\Psi_k\rangle = E_k |\Psi_k\rangle, \ k = 0, 1, 2, \ldots, d-1.
\end{eqnarray} 
When the dimension $d$ becomes prohibitive for the exact diagonalization method, we can resort to the projection technique, i.e., to project the full-model-space eigenvalue problem into a much smaller model space (called $\mathbbm{P}$-space) eigenvalue problem of an effective Hamiltonian. In this context, given a complete set of basis states, $\{ | \Phi_k \rangle \}$ ($k=0, 1, 2, \ldots ,d-1$), we introduce two projection operators $P$ and $Q$. $P$ projects the full $d$-dimensional model space on a smaller $d_p$-dimensional $\mathbbm{P}$-space,
\begin{eqnarray}
P \equiv \sum_{k \in \mathbbm{P}} | \Phi_k \rangle \langle \Phi_k |
\end{eqnarray}
while $Q$ projects the full model space on the $d_q$-dimensional complementary space (called $\mathbbm{Q}$-space),
\begin{eqnarray}
Q = \sum_{k \in \mathbbm{Q}} | \Phi_k \rangle \langle \Phi_k |, 
\end{eqnarray}
where $d_p+d_q = d$. These projection operators satisfy 
\begin{eqnarray}
P+Q=1, \, P^2=P, \, Q^2 = Q, \, PQ = QP = 0. 
\end{eqnarray}
To reduce the dimensionality of the eigenvalue problem, we project the eigenvalue equation (\ref{eq:full_space_schrodinger_equation}) for $H$ into the small $\mathbbm{P}$-space,
\begin{eqnarray}\label{eq:p_space_schrodinger_equation}
H_{\rm eff}(E_k) |\Psi_k^\mathbbm{P}\rangle = E_k |\Psi_k^\mathbbm{P}\rangle, \ 
k = 0, 1, 2, \ldots, d_p-1, 
\end{eqnarray}
for an energy-dependent {\em effective} Hamiltonian~\cite{BlochHorowitz1958}
\begin{eqnarray}\label{eq:energy_dependent_effective_hamiltonian}
H_{\rm eff}(E_k) \equiv PHP + PHQ \frac{1}{E_k-QHQ} QHP, 
\end{eqnarray}
where $|\Psi_k^\mathbbm{P}\rangle \equiv P |\Psi_k\rangle$ is the $\mathbbm{P}$-space component of the $k$th full model space eigenstate $|\Psi_k\rangle$. The full model space eigenstates of Eq.~(\ref{eq:full_space_schrodinger_equation}) can be restored via 
\begin{eqnarray}
|\Psi_k\rangle = |\Psi_k^\mathbbm{P}\rangle + \frac{1}{E_k-QHQ}QHP |\Psi_k^\mathbbm{P}\rangle, 
\end{eqnarray}
and the corresponding $\mathbbm{Q}$-space component is thus 
\begin{eqnarray}
|\Psi_k^\mathbbm{Q}\rangle \equiv Q |\Psi_k\rangle = \frac{1}{E_k-QHQ}QHP |\Psi_k^\mathbbm{P}\rangle. 
\end{eqnarray}
We note that the effective Hamiltonian depends on the exact eigenenergies $E_k$, and hence the $\mathbbm{P}$-space eigenvalue problem Eq.~(\ref{eq:p_space_schrodinger_equation}) needs to be solved self-consistently. 

Let us define the eigenvalue problem for $H_{\rm eff}(E)$ at an arbitrary energy $E$ as 
\begin{eqnarray}\label{eq:p_space_schrodinger_equation_at_arbitrary_energy}
H_{\rm eff}(E) |\psi_n(E)\rangle = f_n(E) |\psi_n(E)\rangle, \ 
n = 0, 1, \ldots, d_p{-}1. 
\end{eqnarray} 
Here energy $E$ does not necessarily coincide with the eigenenergy and therefore the eigenvalues $f_n(E)$ and eigenvectors $|\psi_n(E)\rangle$ are functions of energy $E$. Remark that each $f_n(E)$ exhibits singularities because of the presence of the resolvent operator $(E-QHQ)^{-1}$ in $H_{\rm eff}(E)$, as shown in Eq.~(\ref{eq:energy_dependent_effective_hamiltonian}). Obviously, the singularities are the eigenvalues of $QHQ$ characterized by the same good quantum numbers (angular momentum, parity, etc) as carried by $f_n(E)$, due to the presence of $PHQ$ and $QHP$ operators in Eq.~(\ref{eq:energy_dependent_effective_hamiltonian}). The eigenvalues $E_k$ of Eq.~(\ref{eq:p_space_schrodinger_equation}) are the energies satisfying $f_n(E)=E$. In practice these energies can be found by finding the intersections of $y=E$ and $y=f_n(E)$, which can be quickly located with the Newton-Raphson method. At each intersection $E=E_k$, only one of the $d_p$ eigenpairs of Eq.~(\ref{eq:p_space_schrodinger_equation_at_arbitrary_energy}) satisfies $f_n(E_k)=E_k$ and $|\psi_n(E_k)\rangle=|\Psi_k^\mathbbm{P}\rangle$, assuming that the eigenenergies of Eq.~(\ref{eq:p_space_schrodinger_equation}) are nondegenerate. Examples with one-dimensional $\mathbbm{P}$-space can be found in Ref.~\cite{BWMBPT2023} and are also given in Fig.~\ref{fig:Li5_DJ16_hw18_Nm2_HO_matrix_inversion} of this paper. The cases with multi-dimensional $\mathbbm{P}$-space are given in Figs.~\ref{fig:Li6_DJ16_hw18_Nm2_HO_matrix_inversion},~\ref{fig:Li7_DJ16_hw18_Nm2_HO_Exact},~and~\ref{fig:Li7_N3LObare_hw18_Nm2_HO_Exact}.

The first derivative of $f_n(E)$ can be easily obtained 
\begin{eqnarray}
f_n'(E) 
= - \frac{\langle\psi_n(E)| PHQ \frac{1}{(E-QHQ)^2} QHP |\psi_n(E)\rangle}
{\langle\psi_n(E)| \psi_n(E)\rangle}.
\end{eqnarray}
We note that $f_n'(E)\leq 0$ for all $E$. At the intersections, i.e., $f_n(E_k)=E_k$, $|\psi_n(E_k)\rangle=|\Psi_k^\mathbbm{P}\rangle$, we have 
\begin{eqnarray}
f_n'(E_k) 
= - \frac{\langle\Psi_k^\mathbbm{P}| PHQ \frac{1}{(E_k-QHQ)^2} QHP |\Psi_k^\mathbbm{P}\rangle}
{\langle\Psi_k^\mathbbm{P}|\Psi_k^\mathbbm{P}\rangle}
\leq 0, 
\end{eqnarray}
which is related to the occupation probability ratio of $\mathbbm{Q}$-space to $\mathbbm{P}$-space components of the $k$-th eigenstate of Eq.~(\ref{eq:p_space_schrodinger_equation}) via 
\begin{eqnarray}
\frac{\langle\Psi_k|Q|\Psi_k\rangle}{\langle\Psi_k|P|\Psi_k\rangle}
&=& \frac{\langle\Psi_k^\mathbbm{P}|PHQ\frac{1}{(E_k-QHQ)^2}QHP|\Psi_k^\mathbbm{P}\rangle}
{\langle\Psi_k^\mathbbm{P}|\Psi_k^\mathbbm{P}\rangle} \nonumber\\
&=& - f_n'(E_k) \nonumber\\
&\geq& 0. 
\end{eqnarray}
It follows that for $\mathbbm{P}$-space dominated eigenstates, we have $|f_n'(E_k)|<1$, while for $\mathbbm{Q}$-space dominated eigenstates we have $|f_n'(E_k)|>1$.

Because of the energy-dependence of the effective Hamiltonian, all the eigenenergies $E_k$ (carried by the same good quantum numbers as carried by the $\mathbbm{P}$-space) of the full-model-space eigenvalue problem Eq.~(\ref{eq:full_space_schrodinger_equation}) can be found with the method described above. The corresponding $\mathbbm{P}$-space eigenvectors $|\Psi_k^\mathbbm{P}\rangle$ for these eigenenergies are non-orthogonal, since they are the eigenstates of the Hermitian effective Hamiltonian $H_{\rm eff}(E)$ at different energies.  

Until now, we have not mentioned the way to construct the effective Hamiltonian $H_{\rm eff}(E)$. The straightforward and exact way to construct $H_{\rm eff}(E)$ is by taking the inverse of the $(E-QHQ)$ matrix in the $\mathbbm{Q}$-space directly. This matrix inversion method, however, cannot be applied to large model spaces, and hence it serves mainly as a benchmark in this work. In the following subsections, we show how to compute $H_{\rm eff}(E)$ perturbatively within the BW framework.  

\subsection{Convergence Criterion of the Brillouin-Wigner Perturbation Series for the Effective Hamiltonian}\label{sec:convergence_criterion} 

We use in Eq.~(\ref{eq:full_space_schrodinger_equation}) the $A$-body intrinsic Hamiltonian 
\begin{eqnarray}\label{eq:intrinsic_operator}
H &=& \sum_{i=1}^A \left(\frac{\bm{p}_i^2}{2m} + u_i \right) 
+ \sum_{\substack{i<j}}^A \left( V_{ij} - \frac{\bm{p}_i\cdot\bm{p}_j}{mA} \right) \nonumber\\
&& - \sum_{i=1}^A \left( u_i + \frac{\bm{p}_i^2}{2mA} \right),
\end{eqnarray}
where $m$ is the nucleon mass (approximated here as the average of the neutron and proton mass), $\bm{p}_i$ is the $i$th nucleon's momentum, and $V_{ij}$ is the nucleon-nucleon interaction with additional Coulomb interaction for protons. We have introduced an auxiliary potential $u$ so that the eigenvalue problem for the first term $H_0 \equiv \sum_{i=1}^A \left(\bm{p}_i^2/2m + u_i \right)$ (i.e., the unperturbed Hamiltonian) can be easily solved. The rest part of the intrinsic Hamiltonian $H_1\equiv H - H_0$ is the residual interaction. In this work, we use the spherically-symmetric HO potential as the auxiliary potential, i.e., $u=\frac{1}{2}m\omega^2\bm{r}^2$, where $\omega$ is the HO angular frequency. We denote the eigenvalue equation of $H_0$ as 
\begin{eqnarray}
H_0 |\Phi_k\rangle = \mathcal{E}_k |\Phi_k\rangle, 
\end{eqnarray}
where $\mathcal{E}_k$ are the sum of $A$ single-particle HO energies, and $|\Phi_k\rangle$ are the $A$-body HO Slater determinants. In this work, the full model space is spanned by the $M$-scheme HO Slater determinants (characterized by good quantum numbers: $z$-component of total angular momentum $M$, parity $\pi$ and $z$-component of total isospin $M_T$) and truncated by $N_{\rm max}$ (the total HO excitation quantum above the lowest configuration), as commonly used in NCSM calculations~\cite{Barrett2013}. We choose the $M$-scheme basis state(s) with the lowest energy (the so-called $0\hbar\omega$ model space) as the $\mathbbm{P}$-space. Therefore the $\mathbbm{P}$-space is one-dimensional for closed-shell nuclei and multi-dimensional for open-shell nuclei. 

Now let us consider the exact resolvent operator $(E-QHQ)^{-1}$ in $H_{\rm eff}(E)$ and  introduce an additional partitioning parameter $\xi $, which is an operator, diagonal in the $\mathbbm{Q}$-space. The resolvent operator can be expanded to a BW perturbation series 
\begin{eqnarray}\label{eq:exact_resolvent_operator_expansion}
&& \frac{1}{E - QHQ} = \frac{1}{E - QH_0Q - QH_1Q} \nonumber\\
&=& \frac{1}{\underbrace{(E- QH_0Q-Q\xi Q)}_{X} - \underbrace{(QH_1Q-Q\xi Q)}_{Y}} \nonumber\\
&=& \frac{1}{X} + \frac{1}{X} Y \frac{1}{X} + \frac{1}{X} Y \frac{1}{X} Y \frac{1}{X} + \frac{1}{X} Y \frac{1}{X} Y \frac{1}{X} Y \frac{1}{X} + \cdots \nonumber\\
&=& \lim_{n\rightarrow\infty}\sum_{k=0}^n R^k \frac{1}{X}, 
\end{eqnarray}
where 
\begin{eqnarray}
R \equiv \frac{1}{X} Y = \frac{1}{E-QH_0Q-Q\xi Q} (QH_1Q-Q\xi Q)
\end{eqnarray}
is the expansion ratio depending on the energy $E$, and $X$ is diagonal in the $\mathbbm{Q}$-space so that its inverse can be easily calculated. In the derivation of Eq.~(\ref{eq:exact_resolvent_operator_expansion}), we repeatedly used the following identity:
\begin{eqnarray}\label{eq:x_y_identity}
\frac{1}{X-Y} 
= \frac{1}{X} + \frac{1}{X-Y} Y \frac{1}{X}. 
\end{eqnarray}
Inserting Eq.~(\ref{eq:exact_resolvent_operator_expansion}) back to the effective Hamiltonian $H_{\rm eff}(E)$, we are able to construct $H_{\rm eff}(E)$ perturbatively, providing the BW perturbation series is converging. The BW series in Eq.~(\ref{eq:exact_resolvent_operator_expansion}) is nothing but an operator-valued geometric series, which converges to the exact resolvent operator if and only if the spectral radius (the maximum absolute eigenvalues) of $R$ is smaller than unity, i.e., $\rho(R)<1$~\cite{Lowdin1962,Horn2012matrix}. We refer to this condition as the \textit{convergence criterion} of BW perturbation series~\cite{BWMBPT2023}. Noticeably, the smaller the value of $\rho(R)$ is, the faster the speed of convergence will be. The partitioning parameter $\xi$ we introduced here can be used to tune the convergence behavior of the BW series. 

Note that the expansion ratio $R$ depends on the energy $E$. Let $E_{\rm min}^{qhq}$ and $E_{\rm max}^{qhq}$ be the lowest and the highest eigenvalues of the operator $QHQ$ in the $\mathbbm{Q}$-space. As we have shown in Ref.~\cite{BWMBPT2023}, for energy $E<E_{\rm min}^{qhq}$ the convergence criterion $\rho(R)<1$ for $f_n(E)$ can always be satisfied with a specific choice of the partitioning parameter $\xi$. We refer to this energy interval $-\infty<E<E_{\rm min}^{qhq}$ as the convergence interval. This convergence interval was also discussed in Refs.~\cite{Schucan1972,Schucan1973} in the context of the convergence of a Rayleigh-Schr\"odinger perturbative calculation. Recall that $H_0$ and $\xi$ are diagonal operators in the $\mathbbm{Q}$-space. We denote their diagonal entries as $\{\mathcal{E}_1^q \leq \mathcal{E}_2^q \leq \mathcal{E}_3^q \leq \cdots \leq \mathcal{E}_{d_q}^{q}\}$ and $\{\xi_1^q \leq \xi_2^q \leq \xi_3^q \leq \cdots \leq \xi_{d_q}^{q}\}$, respectively. We have shown in the supplemental material of Ref.~\cite{BWMBPT2023} that the convergence criterion $\rho(R)<1$ for $E<E_{\rm min}^{qhq}$ can be always satisfied by adjusting $\xi_{k}^q$ ($k = 1,2, \ldots, d_q$) such that
\begin{equation}\label{eq:optimized_xi} 
\mathcal{E}_k^q+\xi_k^q > \dfrac12 \left(E_{\rm min}^{qhq} + E_{\rm max}^{qhq}\right),\ k = 1,2, \ldots, d_q.
\end{equation}
This means that the diagonal entries $\xi_k^q$ should be large enough to satisfy the above condition. Note that the above inequality (\ref{eq:optimized_xi}) is just a sufficient condition and does not necessarily cover all values of $\xi$ which make the perturbation series converging. 

Inserting the BW perturbation series Eq.~(\ref{eq:exact_resolvent_operator_expansion}) back into the effective Hamiltonian $H_{\rm eff}(E)$, we have to consider the symmetries preserved by the intrinsic Hamiltonian $H$ because of the presence of the $PHQ$ and $QHP$ operators in $H_{\rm eff}(E)$, as shown in Eq.~(\ref{eq:energy_dependent_effective_hamiltonian}). Therefore, the $\mathbbm{Q}$-space and its projection operator $Q$ used in the above analysis should be replaced with its subspace $\mathbbm{Q}_s$ and the corresponding projection operator $Q_s$, characterized by the same good quantum numbers (e.g. angular momentum $J$, parity $\pi$, center-of-mass quantum numbers $N_{\rm cm}$, $L_{\rm cm}$, $M_{\rm cm}$, etc) as the eigenvalues $f_n(E)$. Therefore $E_{\rm min}^{qhq}$ becomes the lowest eigenvalue of the $Q_sHQ_s$ operator, which is exactly the lowest singularity of $f_n(E)$. Note that we choose the $0\hbar\omega$ model space as our $\mathbbm{P}$-space, whose center-of-mass motion is in the ground state, i.e., $N_{\rm cm}=L_{\rm cm}=M_{\rm cm}=0$. Therefore, the subspace $\mathbbm{Q}_s$ does not contain spurious center-of-mass motions. 

In Ref.~\cite{BWMBPT2023} we concluded that for ground states of closed-shell nuclei, the convergence criterion $\rho(R)<1$ is always satisfied by choosing large enough $\xi$ due to the variational principle. For open-shell nuclei, the $0\hbar\omega$ model space is not one-dimensional anymore, and hence excited states can also be calculated. The conclusion for the ground states of closed-shell nuclei can now be generalized to the lowest $J^\pi$ states for open-shell nuclei, i.e., the convergence criterion can always be satisfied for each lowest $J^\pi$ state by choosing large enough $\xi$ (namely large enough diagonal entries of $\xi$) due to the variational principle and the conserved quantum numbers $J$ and $\pi$. For excited states of each $J^\pi$, only those states with eigenenergies in the convergence interval $E_k<E_{\rm min}^{qhq}$ can get converged. The presence of intruder states\footnote{$\mathbbm{Q}$-space dominated states which occur in the energy region of $\mathbbm{P}$-space dominated states.} may reduce the size of the convergence interval and spoil the convergence of higher excited states of each $J^\pi$. We will discuss this with numerical results in Sec.~\ref{sec:results}.  

\subsection{$\hat{K}$-Box Iterative Calculation of High-Order Terms of Brillouin-Wigner Perturbation Series}

Before going into the numerical calculations, we here outline the $\hat{K}$-box iterative method introduced in Ref.~\cite{BWMBPT2023} to efficiently calculate the high order terms of the BW perturbation series for the effective Hamiltonian $H_{\rm eff}(E)$. Applying Eq.~(\ref{eq:x_y_identity}) into the exact resolvent operator in $H_{\rm eff}(E)$ we obtain 
\begin{eqnarray} 
H_{\rm eff}(E) 
= PHP + P\hat{K}(E)Q \frac{1}{E-QH_0Q-Q\xi Q} QHP, \nonumber\\
\end{eqnarray}
where we have defined a special vertex function $\hat{K}$-box in $\mathbbm{P}\mathbbm{Q}$-space, introduced first in Ref.~\cite{BWMBPT2023}~: 
\begin{eqnarray}
\hat{K}(E) \equiv PHQ + PHQ \frac{1}{E-QHQ} (QH_1Q-Q\xi Q). 
\end{eqnarray}
Applying again Eq.~(\ref{eq:x_y_identity}) to the resolvent operator in the above equation,
we obtain the following recursive equation for $\hat{K}(E)$:
\begin{eqnarray}
\hat{K}(E) 
= PHQ + P\hat{K}(E)Q \frac{1}{E-(H_0+\xi)} Q(H_1-\xi)Q. \nonumber\\ 
\end{eqnarray}
The value of $\hat{K}(E)$ can be calculated iteratively via
\begin{eqnarray}\label{eq:kbox_iterative_equation}
\hat{K}^{(s)}(E) 
&=& PHQ \nonumber\\
&& + P\hat{K}^{(s-1)}(E)Q \frac{1}{E-(H_0+\xi)} Q(H_1-\xi)Q, \nonumber\\
\end{eqnarray}
starting with $\hat{K}^{(0)}(E)=0$, where $s=1, 2, 3, \cdots$. The $(s-1)$th iterative result $\hat{K}^{(s-1)}(E)$ corresponds to the accumulated BW perturbation series of $H_{\rm eff}(E)$ up to the $s$th order, $H_{\rm eff}^{s{\rm th}}(E)$. We denote the corresponding eigenvalues as $f_n^{s{\rm th}}(E)$. With the above iterative calculations of $\hat{K}$-box, we can easily reach high orders of BW perturbation series of $H_{\rm eff}(E)$. The advantage of using $\hat{K}$-box is that we only need to store a $(d_p\times d_q)$ matrix $\hat{K}$ in the memory during iterations. 

The full model space eigenvectors can be calculated from $\hat{K}$-box via  
\begin{eqnarray}
|\Psi_k\rangle 
= |\Psi_k^\mathbbm{P}\rangle 
+ \frac{1}{E_k-QH_0Q-Q\xi Q} \hat{K}^\dag(E_k) |\Psi_k^\mathbbm{P}\rangle.
\end{eqnarray}
where we made use of another equality, namely,
\begin{eqnarray}\label{eq:x_y_identity2}
\frac{1}{X-Y} 
= \frac{1}{X} + \frac{1}{X} Y \frac{1}{X-Y}. 
\end{eqnarray}
Therefore other physical observables apart from energies can also be calculated with the full-model-space eigenvectors $|\Psi_k\rangle$. 

\begin{figure}[t]
\centering
\subfloat[\label{subfig:Li5_DJ16_exact}]{%
\includegraphics[width=0.5\textwidth]{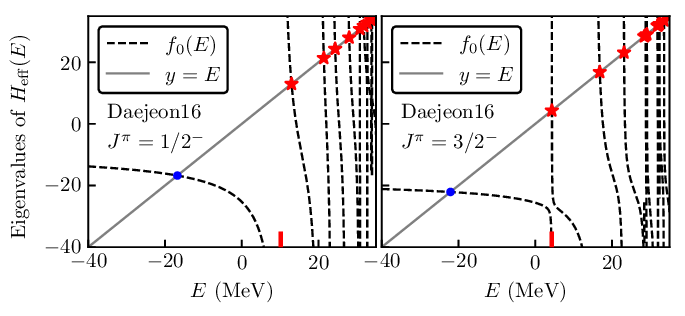}
}\hfill
\subfloat[\label{subfig:Li5_N3LObare_exact}]{%
\includegraphics[width=0.5\textwidth]{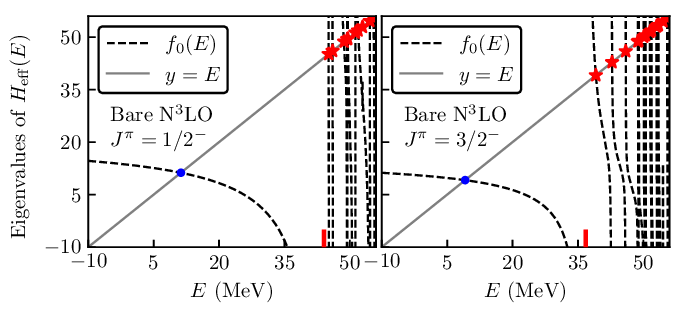}
}\hfill
\caption{\label{fig:Li5_DJ16_hw18_Nm2_HO_matrix_inversion}The eigenvalue functions $f_0(E)$ ($J^\pi = 1/2^-, 3/2^-$) of $H_{\rm eff}(E)$ for $^5$Li calculated from the matrix inversion method using (a) the Daejeon16 potential and (b) the bare N$^3$LO potential with $\hbar\omega=18$~MeV at $N_{\rm max}=2$. The blue dots and red stars are the solutions of Eq.~(\ref{eq:p_space_schrodinger_equation}), which can be exactly reproduced by NCSM. The bottom red vertical lines mark the positions of the lowest eigenvalues of $QHQ$ with $J^\pi = 1/2^-, 3/2^-$ (zero CM excitation), which are the lowest singularities of $f_0(E)$ at $J^\pi = 1/2^-, 3/2^-$, respectively.}
\end{figure}
\begin{figure}[t]
\centering
\includegraphics[width=0.48\textwidth]{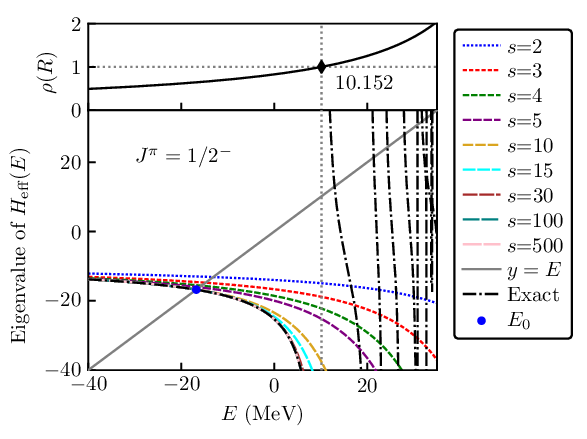} 
\includegraphics[width=0.48\textwidth]{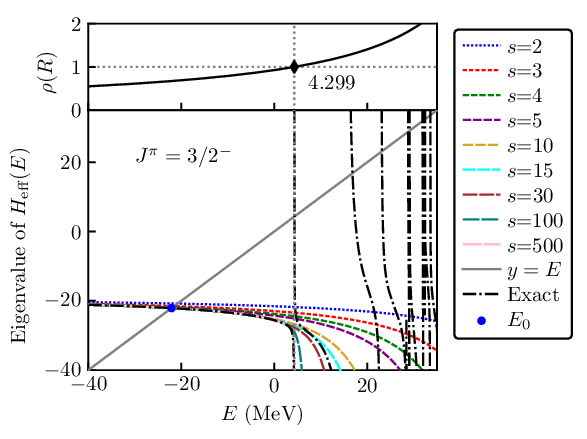} 
\caption{\label{fig:Li5_DJ16_hw18_Nm2_HO_Part1_Kbox_Nmax2}The eigenvalue functions $f_0^{s{\rm th}}(E)$ ($J^\pi = 1/2^-, 3/2^-$) of $H_{\rm eff}^{s{\rm th}}(E)$ from BW perturbative calculations up to various orders $s$ for $^5$Li using the Daejeon16 potential with $\xi = \langle{\rm C}|H_1|{\rm C}\rangle=-130.227$~MeV at $\hbar\omega=18$~MeV, $N_{\rm max}=2$. Inside each panel, the value of $\rho(R)$ with the same $J^\pi$ is depicted on the top, and the exact $f_0(E)$ obtained from matrix inversion is plotted with black dash-dotted lines. The blue dots mark the exact result from NCSM calculations below the lowest singularities.}
\end{figure}
\begin{figure*}[t]
\centering
\subfloat[\label{subfig:Li5_DJ16_Kbox_Nmax246}]{%
\includegraphics[width=0.5\textwidth]{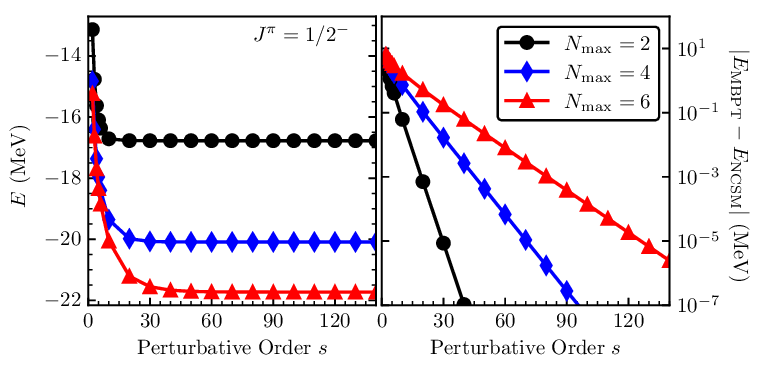} 
\includegraphics[width=0.5\textwidth]{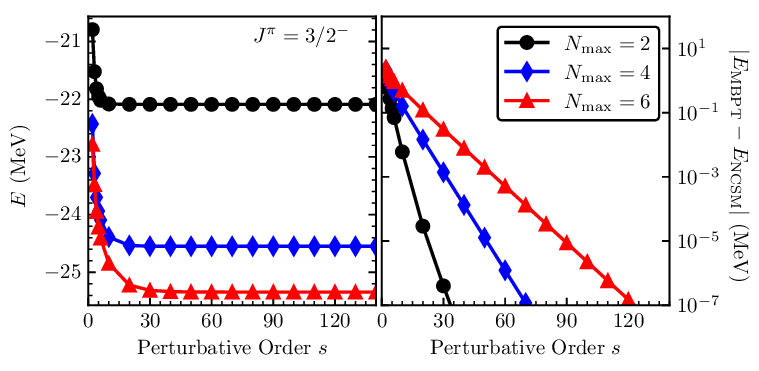} 
}\hfill
\subfloat[\label{subfig:Li5_N3LObare_Kbox_Nmax246}]{%
\includegraphics[width=0.5\textwidth]{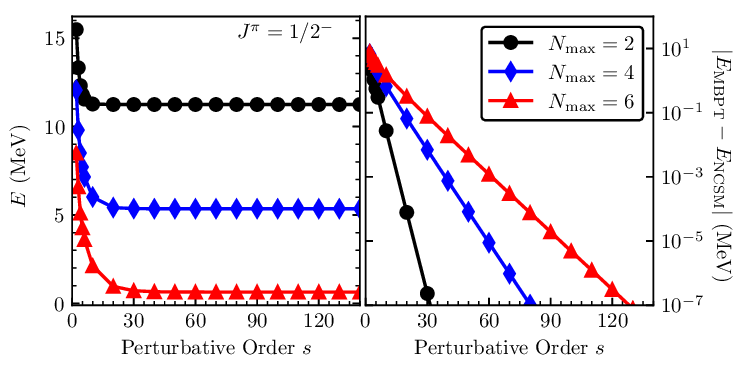} 
\includegraphics[width=0.5\textwidth]{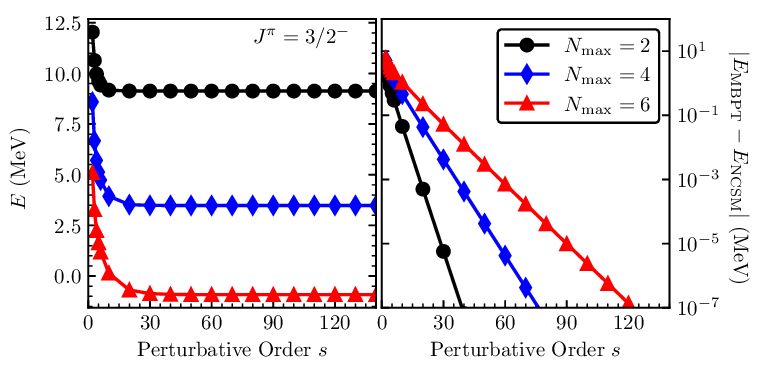} 
}\hfill
\caption{\label{fig:Li5_DJ16_hw18_Nm2_HO_Kbox_Nmax246}The two lowest energies ($J^\pi = 1/2^-, 3/2^-$) of $^5$Li from the BW perturbative calculation up to various orders $s$ using (a) the Daejeon16 potential with $\xi = \langle{\rm C}|H_1|{\rm C}\rangle=-130.227$~MeV and (b) the bare N$^3$LO potential with $\xi = \langle{\rm C}|H_1|{\rm C}\rangle=-103.136$~MeV, at $\hbar\omega=18$~MeV, $N_{\rm max}=2, 4, 6$. Inside each panel, the energy difference in absolute value between the BW perturbative calculation and the NCSM calculation is also plotted.}
\end{figure*}
\begin{figure*}[t]
\centering
\includegraphics[width=0.45\textwidth]{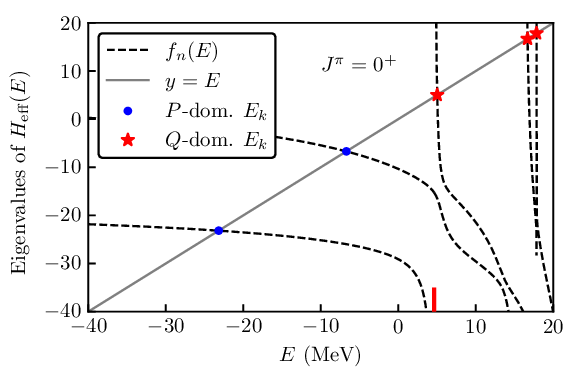} 
\includegraphics[width=0.45\textwidth]{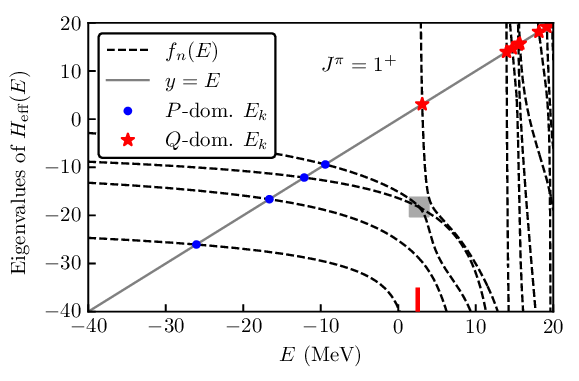} 
\includegraphics[width=0.45\textwidth]{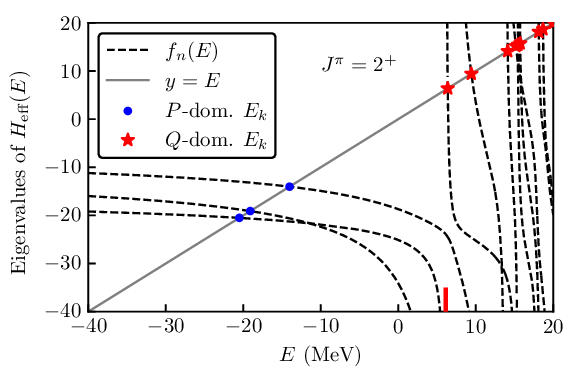} 
\includegraphics[width=0.45\textwidth]{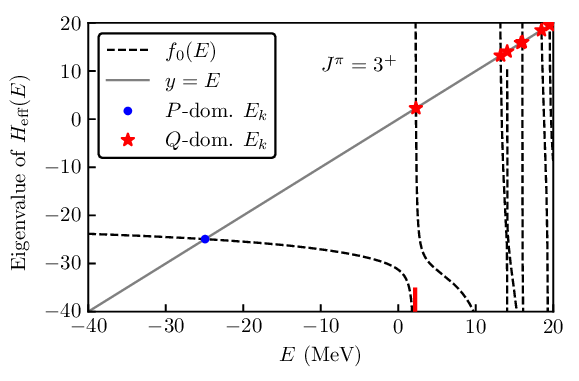} 
\caption{\label{fig:Li6_DJ16_hw18_Nm2_HO_matrix_inversion}The eigenvalue functions $f_n(E)$ ($J^\pi = 0^+, 1^+, 2^+, 3^+$) of $H_{\rm eff}(E)$ for $^6$Li calculated from the exact matrix inversion method using the Daejeon16 potential at $\hbar\omega=18$~MeV, $N_{\rm max}=2$. The blue dots ($\mathbbm{P}$-space dominated) and red stars ($\mathbbm{Q}$-space dominated) are the results from NCSM calculations, which are also the intersections of $y=f_n(E)$ and $y=E$ (i.e., solutions of the $\mathbbm{P}$-space eigenvalue problem Eq.~(\ref{eq:p_space_schrodinger_equation})). The bottom vertical lines mark the position of the lowest eigenvalues of $QHQ$ with $J^\pi = 0^+, 1^+, 2^+, 3^+$ and zero CM excitation, which are the lowest singularities of $f_n(E)$ at $J^\pi = 0^+, 1^+, 2^+, 3^+$ respectively. The details of the eigenvalue ``crossing'' marked by the gray rectangle shadow in the panel of $J^\pi=1^+$ is plotted in Fig.~\ref{fig:Li5_DJ16_hw18_Nm2_HO_Exact_crossing_details}.}
\end{figure*}
\begin{figure}[hbt]
\centering
\includegraphics[height=0.3\textwidth]{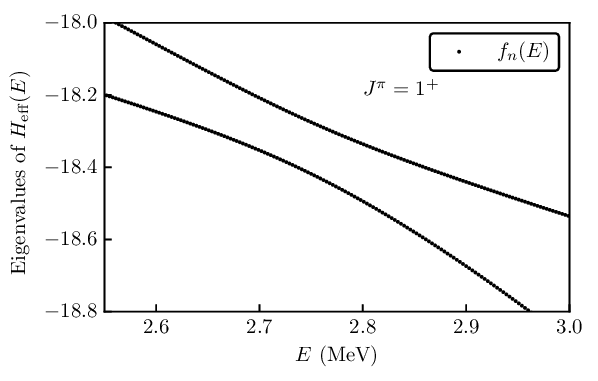} 
\caption{\label{fig:Li5_DJ16_hw18_Nm2_HO_Exact_crossing_details}The details of the eigenvalue ``crossing'' marked by gray rectangle shadow in the $J^\pi=1^+$ panel of Fig.~\ref{fig:Li6_DJ16_hw18_Nm2_HO_matrix_inversion}. The eigenvalue functions $f_n(E)$ avoid crossing.}
\end{figure}
\begin{figure}[hbt]
\centering
\includegraphics[width=0.49\textwidth]{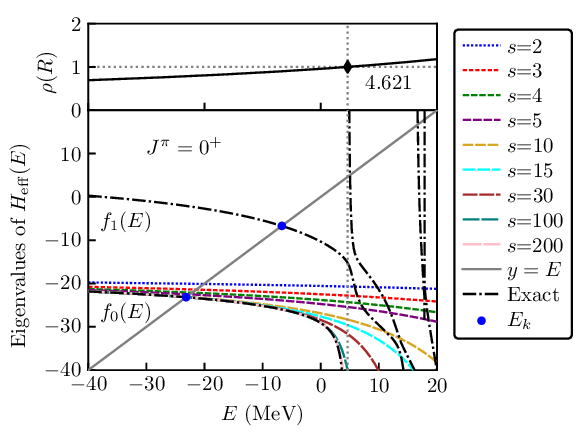} 
\includegraphics[width=0.49\textwidth]{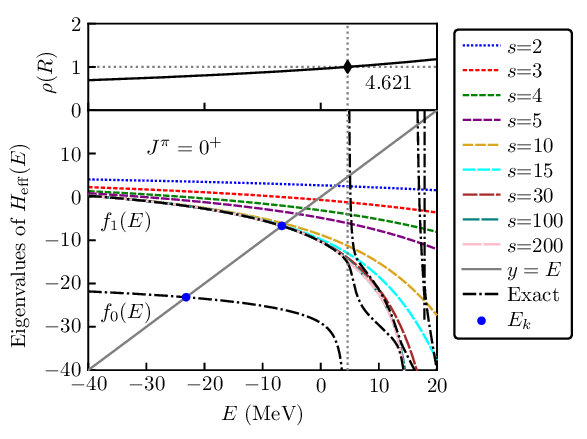}  
\caption{\label{fig:Li6_DJ16_hw18_Nm2_HO_Part1_Kbox_Nmax2}The eigenvalue functions $f_n^{s{\rm th}}(E)$ ($J^\pi=0^+$) of $H_{\rm eff}^{s{\rm th}}(E)$ from the BW perturbative calculation up to various orders $s$ for $^6$Li using the Daejeon16 potential with $\xi = \langle{\rm C}|H_1|{\rm C}\rangle=-128.427$~MeV, at $\hbar\omega=18$~MeV, $N_{\rm max}=2$. For clarity, different states of perturbative calculations are depicted separately. Inside each panel, the value of $\rho(R)$ with the same $J^\pi$ is depicted on the top, and the exact $f_n(E)$ obtained from the matrix inversion method is shown with black dash-dotted lines. The blue dots mark the exact result from NCSM calculations below the lowest singularities.}
\end{figure}
\begin{figure*}[t]
\centering
\includegraphics[width=0.49\textwidth]{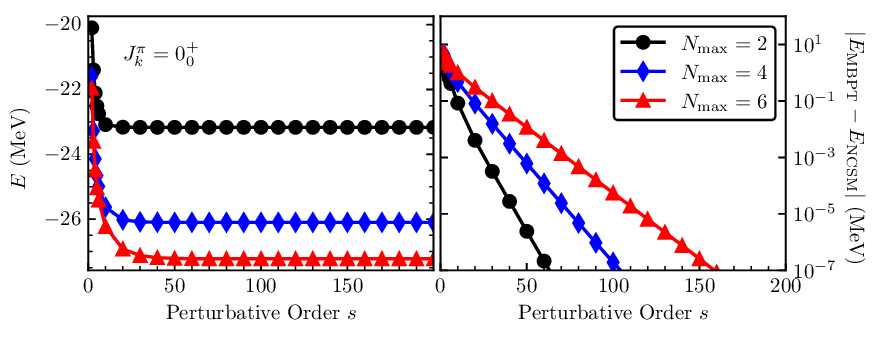} 
\includegraphics[width=0.49\textwidth]{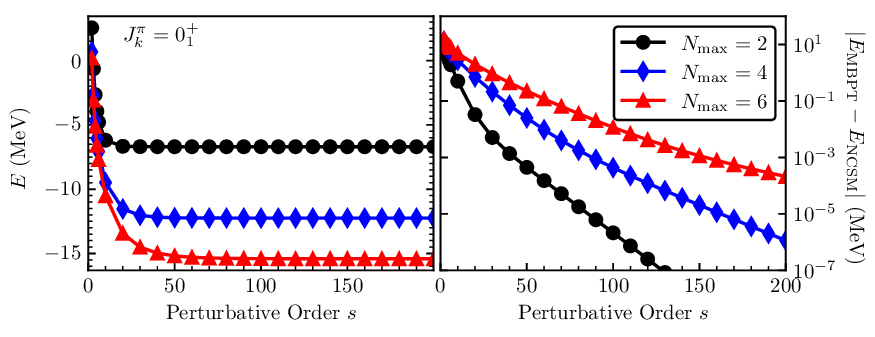} 
\includegraphics[width=0.49\textwidth]{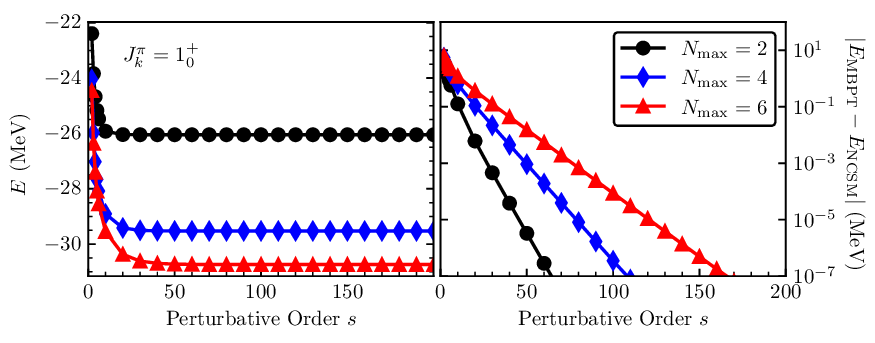} 
\includegraphics[width=0.49\textwidth]{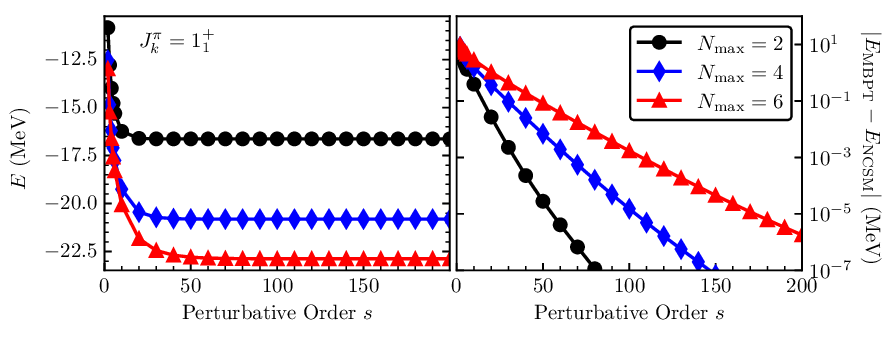} 
\includegraphics[width=0.49\textwidth]{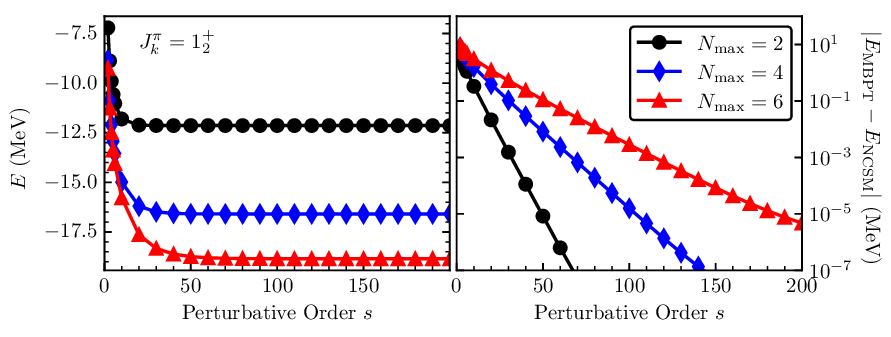} 
\includegraphics[width=0.49\textwidth]{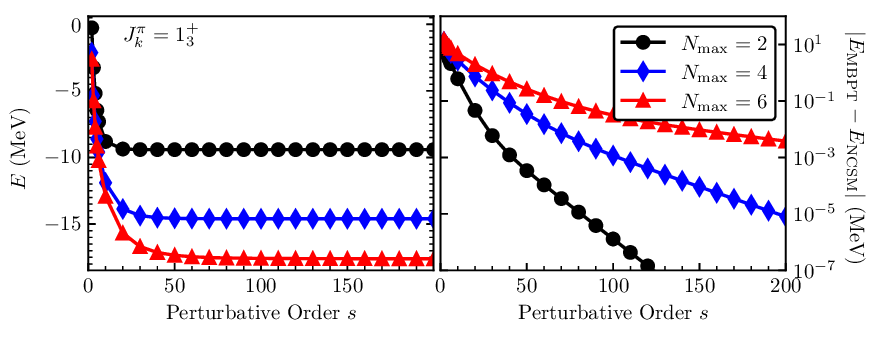} 
\includegraphics[width=0.49\textwidth]{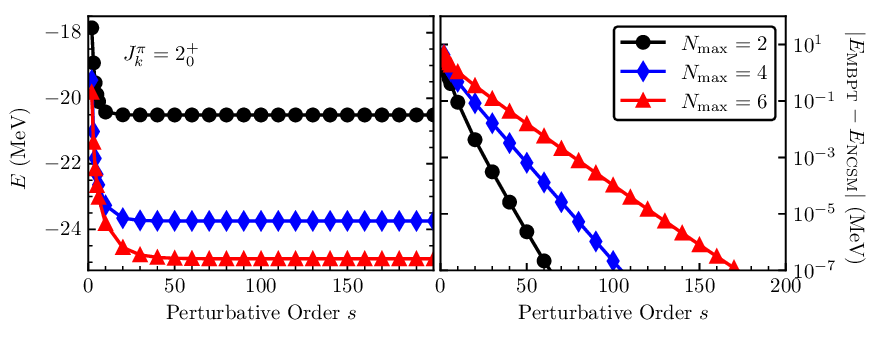} 
\includegraphics[width=0.49\textwidth]{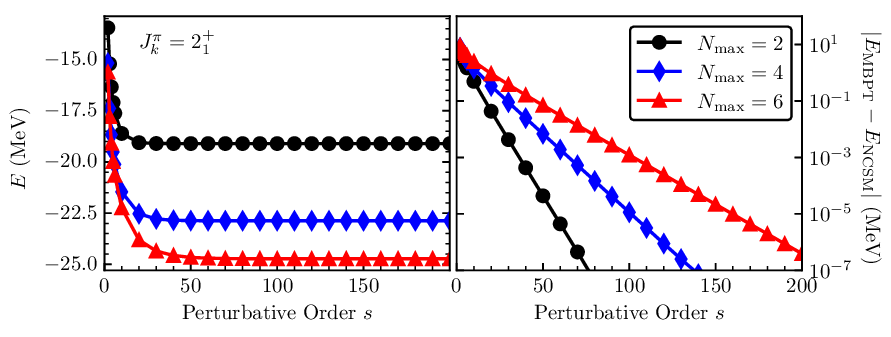} 
\includegraphics[width=0.49\textwidth]{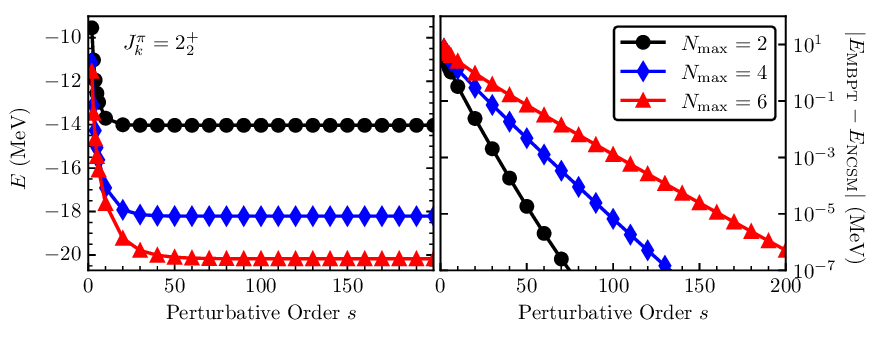} 
\includegraphics[width=0.49\textwidth]{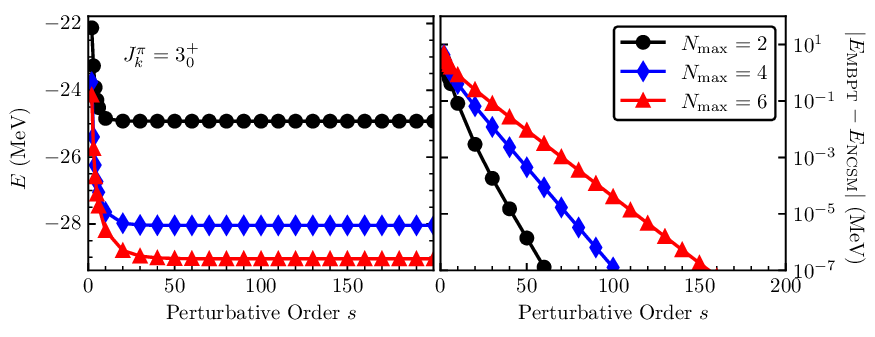} 
\caption{\label{fig:Li_DJ16_hw18_Nm2_HO_Kbox}The energies of low-lying states of $^6$Li ($J^\pi = 0^+, 1^+, 2^+, 3^+$) from the BW perturbative calculation up to various orders $s$ using the Daejeon16 potential with $\xi = \langle{\rm C}|H_1|{\rm C}\rangle=-128.427$~MeV, at $\hbar\omega=18$~MeV, $N_{\rm max}=2, 4, 6$. Inside each panel, the energy difference in absolute value between the BW perturbative calculation and the NCSM calculation is also plotted.}
\end{figure*}
\begin{figure*}[t]
\centering
\includegraphics[width=0.49\textwidth]{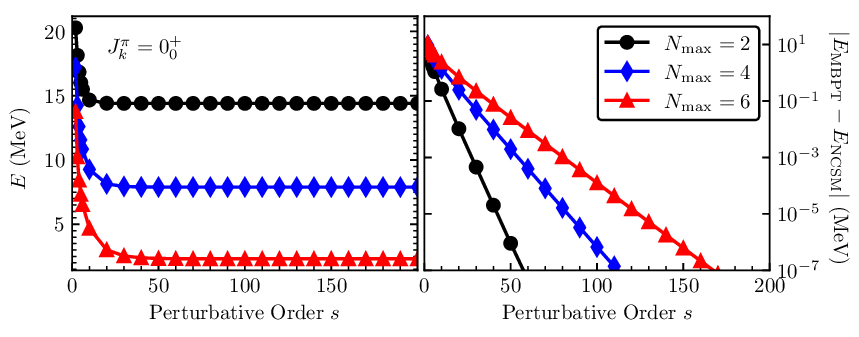} 
\includegraphics[width=0.49\textwidth]{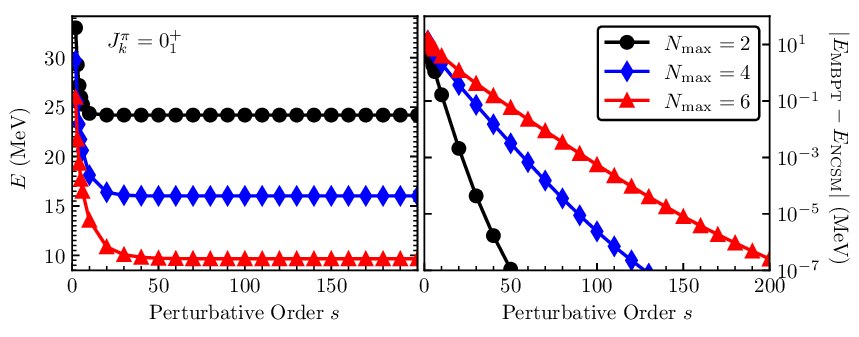} 
\includegraphics[width=0.49\textwidth]{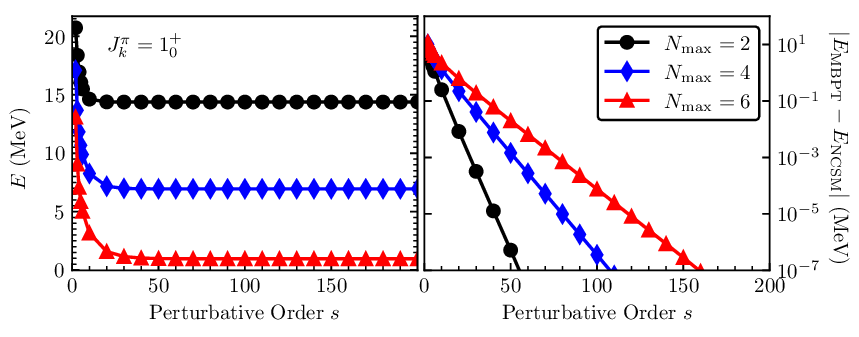} 
\includegraphics[width=0.49\textwidth]{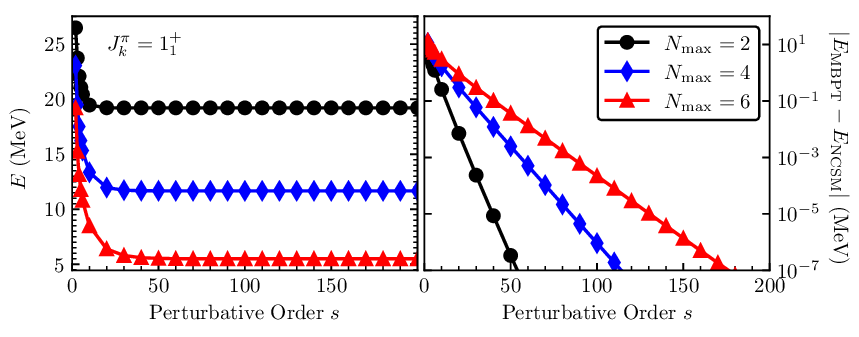} 
\includegraphics[width=0.49\textwidth]{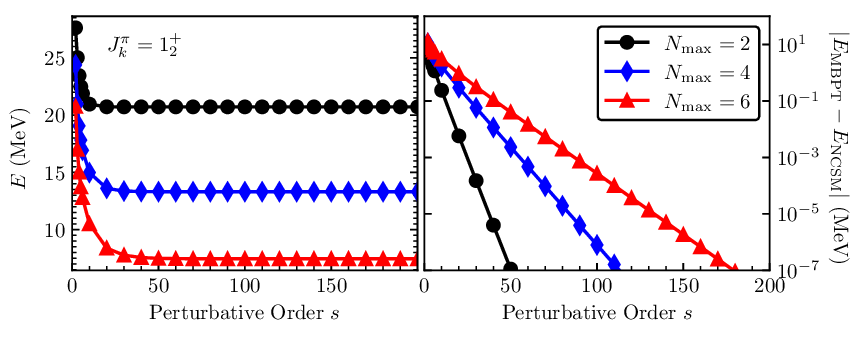} 
\includegraphics[width=0.49\textwidth]{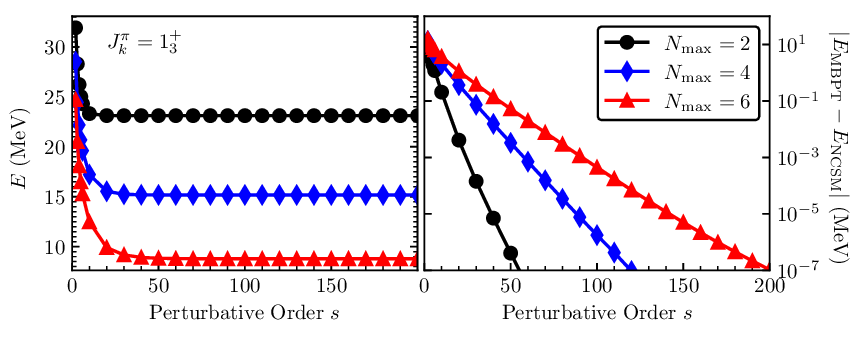} 
\includegraphics[width=0.49\textwidth]{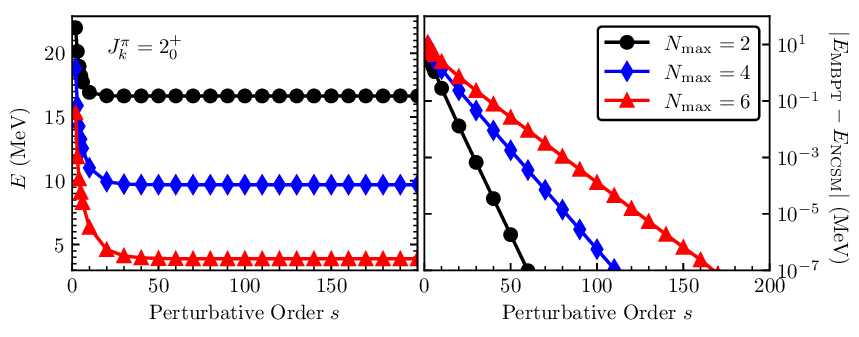} 
\includegraphics[width=0.49\textwidth]{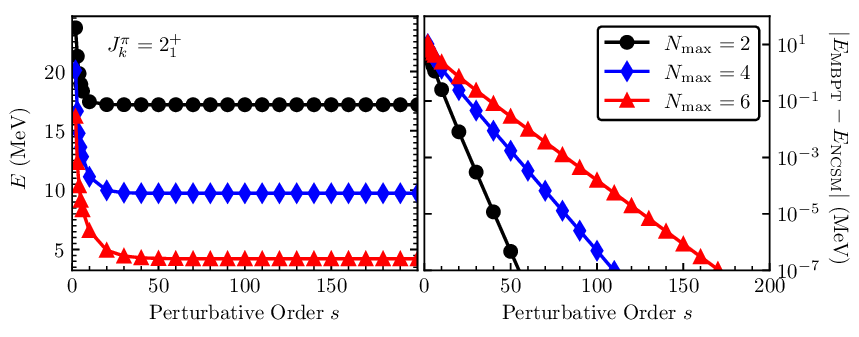} 
\includegraphics[width=0.49\textwidth]{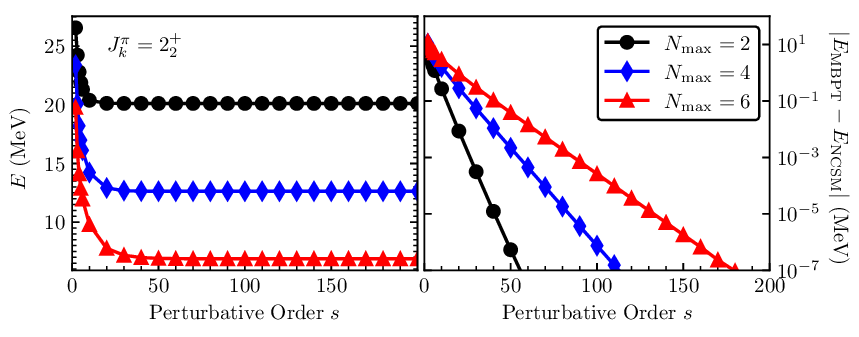} 
\includegraphics[width=0.49\textwidth]{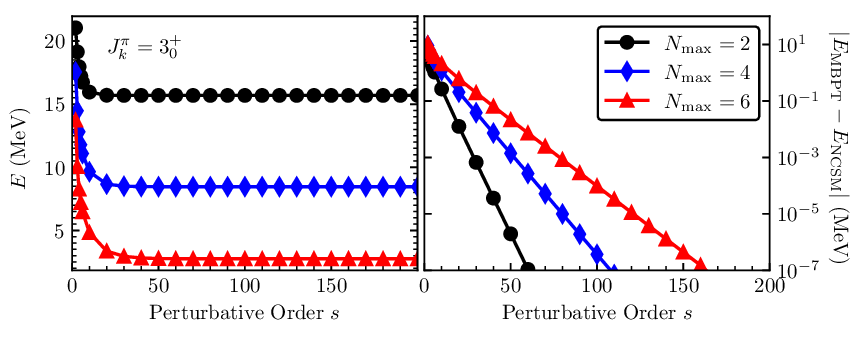} 
\caption{\label{fig:Li6_N3LObare_hw18_Nm2_HO_Kbox}Same as Fig.~\ref{fig:Li_DJ16_hw18_Nm2_HO_Kbox} but with the bare N$^3$LO potential and $\xi = \langle{\rm C}|H_1|{\rm C}\rangle=-101.336$~MeV.}
\end{figure*}
\begin{figure*}[t]
\centering
\includegraphics[width=0.45\textwidth]{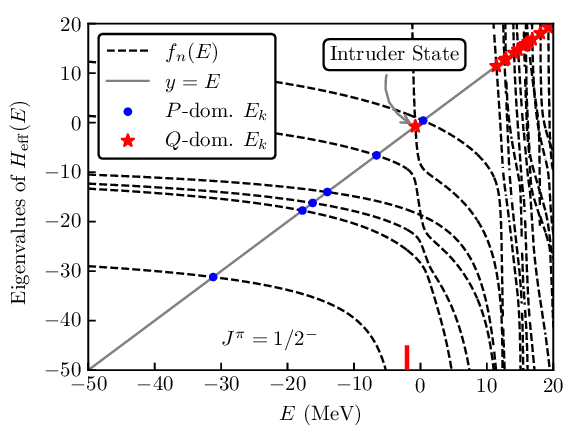} 
\includegraphics[width=0.45\textwidth]{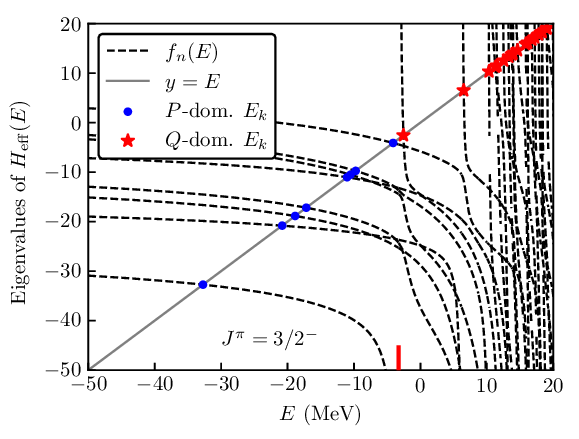} 
\includegraphics[width=0.45\textwidth]{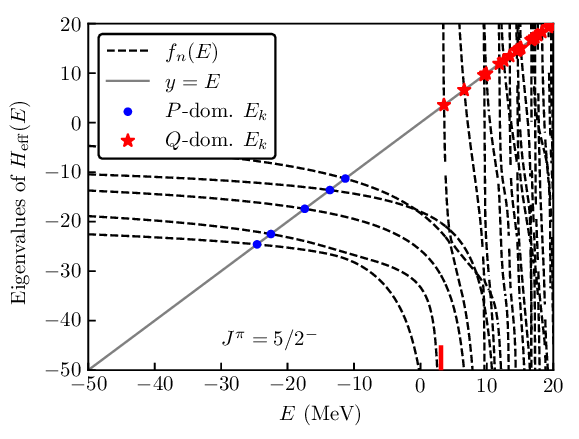} 
\includegraphics[width=0.45\textwidth]{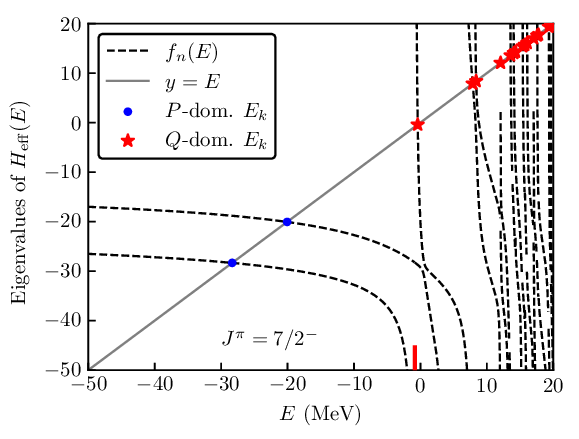} 
\caption{\label{fig:Li7_DJ16_hw18_Nm2_HO_Exact}The eigenvalue functions $f_n(E)$ ($J^\pi = 1/2^-, 3/2^-, 5/2^-, 7/2^-$) of $H_{\rm eff}(E)$ for $^7$Li calculated from the exact matrix inversion method using the Daejeon16 potential at $\hbar\omega=18$~MeV, $N_{\rm max}=2$. The blue dots ($\mathbbm{P}$-space dominated) and red stars ($\mathbbm{Q}$-space dominated) are the eigenenergies of the $\mathbbm{P}$-space eigenvalue problem Eq.~(\ref{eq:p_space_schrodinger_equation}). The red bottom vertical lines mark the position of the lowest eigenvalues of $QHQ$ with corresponding $J^\pi$ and zero CM excitation, which are the lowest singularities of $f_n(E)$.}
\end{figure*}
\begin{figure*}[t]
\centering
\includegraphics[width=0.48\textwidth]{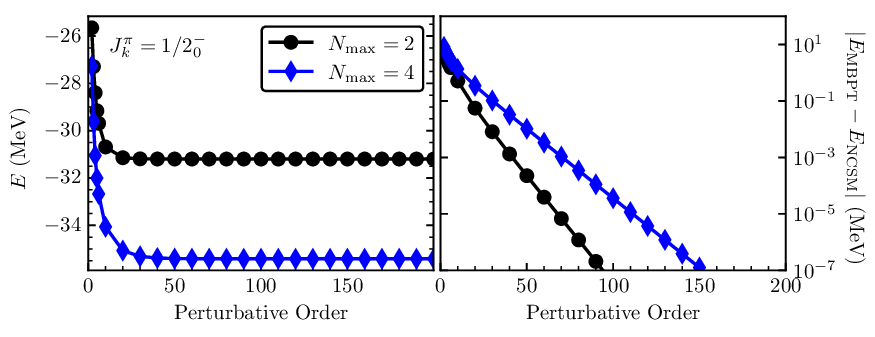} 
\includegraphics[width=0.48\textwidth]{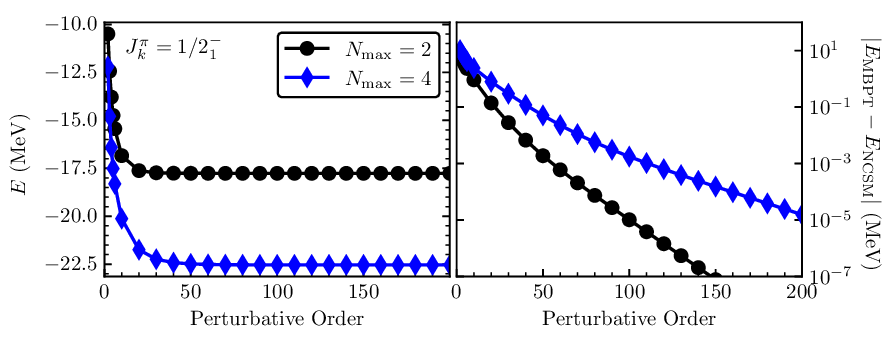} 
\includegraphics[width=0.48\textwidth]{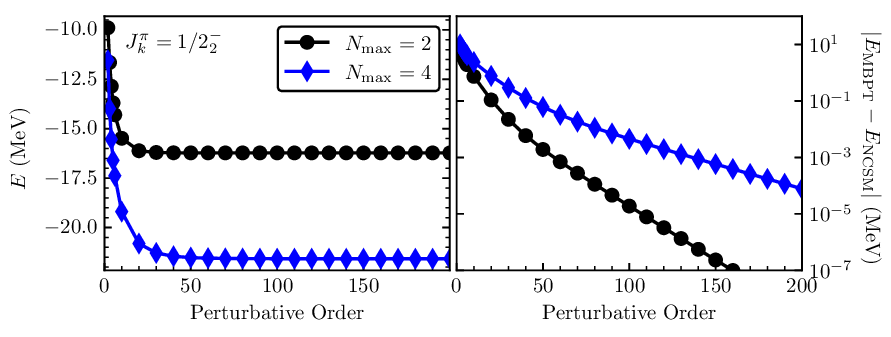} 
\includegraphics[width=0.48\textwidth]{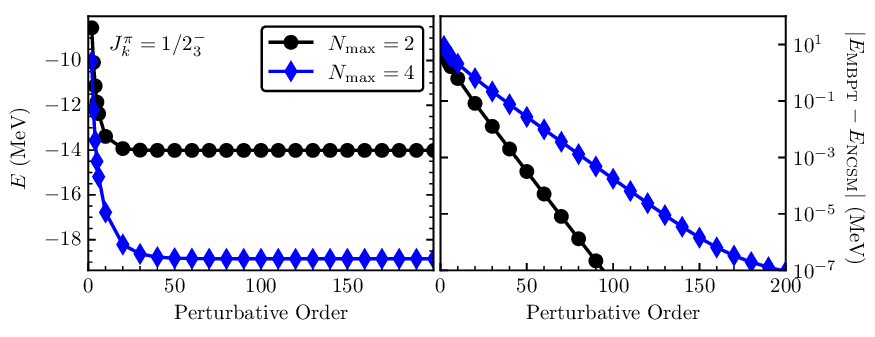} 
\includegraphics[width=0.48\textwidth]{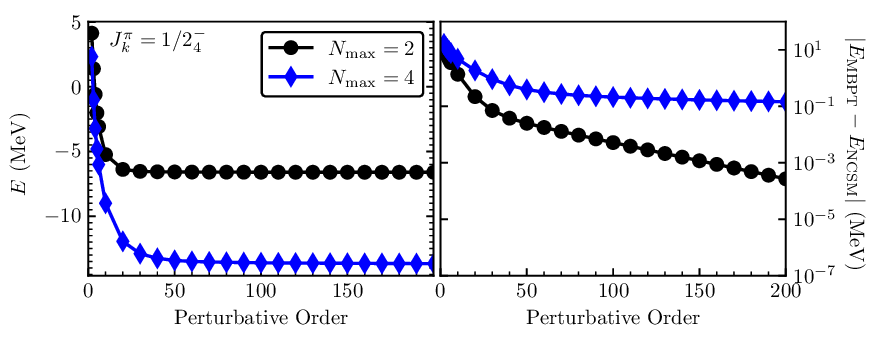} 
\includegraphics[width=0.48\textwidth]{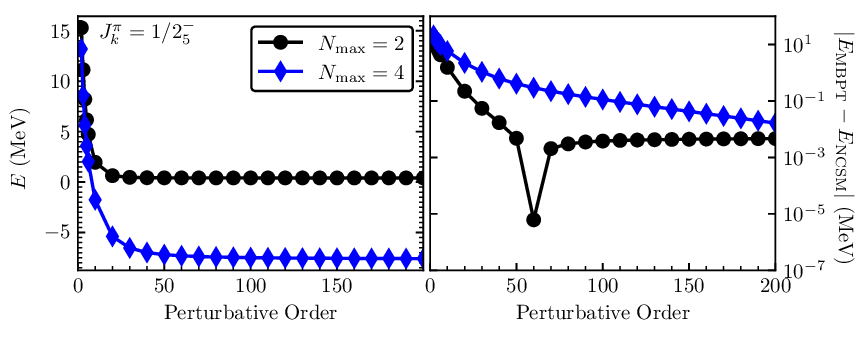} 
\caption{\label{fig:Li7_DJ16_hw18_Nm2_HO_Kbox_Nmax246}The energies of low-lying states of $^7$Li ($J^\pi = 1/2^-$) from the BW perturbative calculation up to various orders using the Daejeon16 potential with $\xi = \langle{\rm C}|H_1|{\rm C}\rangle=-127.141$~MeV, at $\hbar\omega=18$~MeV, $N_{\rm max}=2, 4$. Inside each panel, the energy difference in absolute value between the BW perturbative calculation and the NCSM calculation is also plotted.}
\end{figure*}
\begin{figure*}[hbt]
\centering
\includegraphics[width=0.48\textwidth]{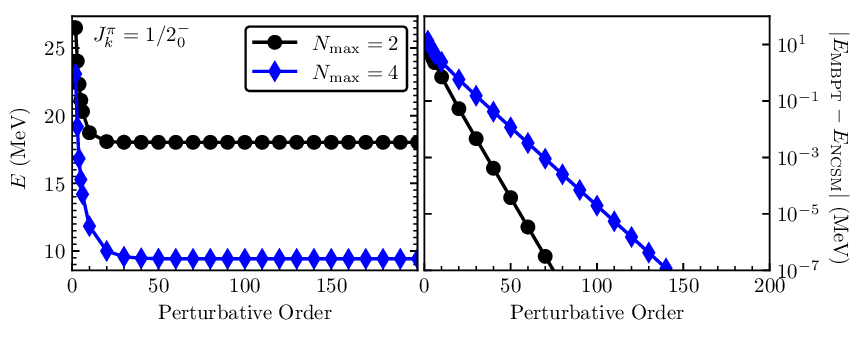} 
\includegraphics[width=0.48\textwidth]{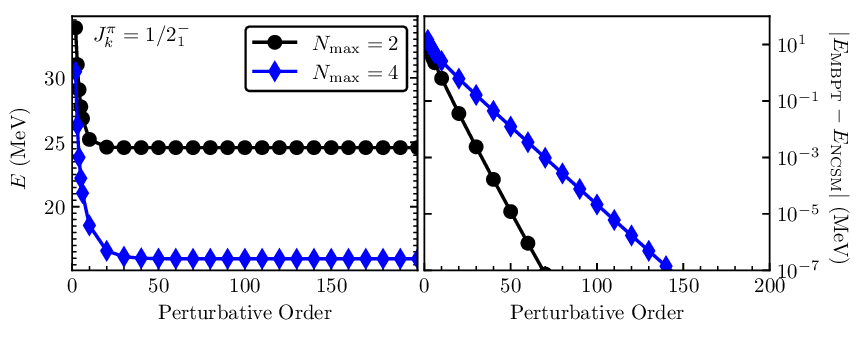} 
\includegraphics[width=0.48\textwidth]{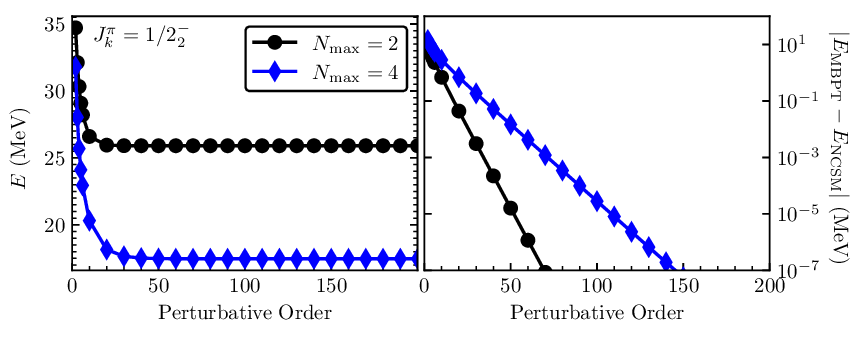} 
\includegraphics[width=0.48\textwidth]{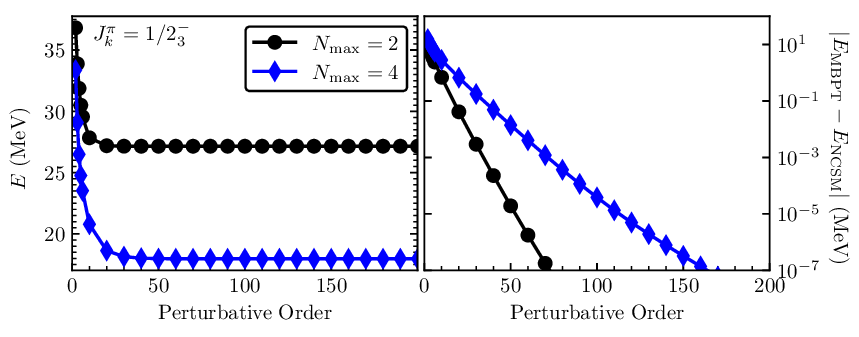} 
\includegraphics[width=0.48\textwidth]{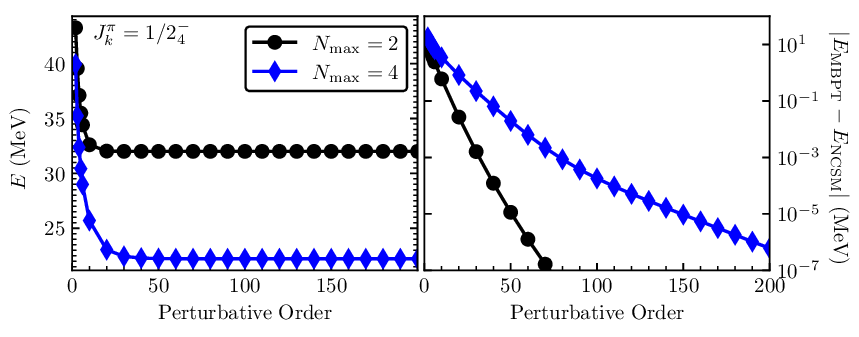} 
\includegraphics[width=0.48\textwidth]{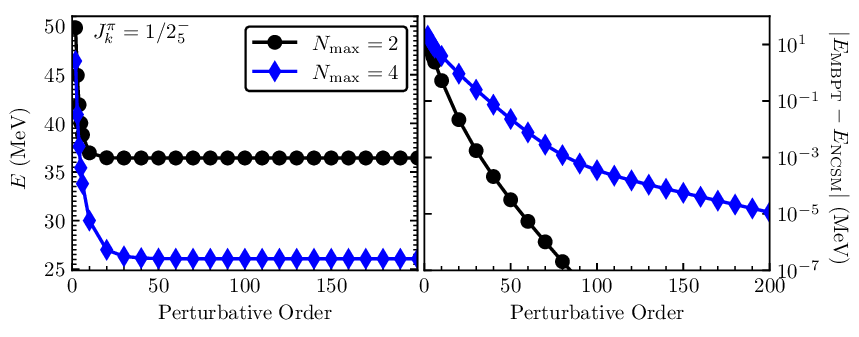} 
\caption{\label{fig:Li7_N3LObare_hw18_Nm2_HO_Kbox_Nmax246}Same as Fig.~\ref{fig:Li7_DJ16_hw18_Nm2_HO_Kbox_Nmax246} but with the bare N$^3$LO potential with $\xi = \langle{\rm C}|H_1|{\rm C}\rangle=-100.050$~MeV}
\end{figure*}
\begin{figure}[t]
\centering
\includegraphics[width=0.4\textwidth]{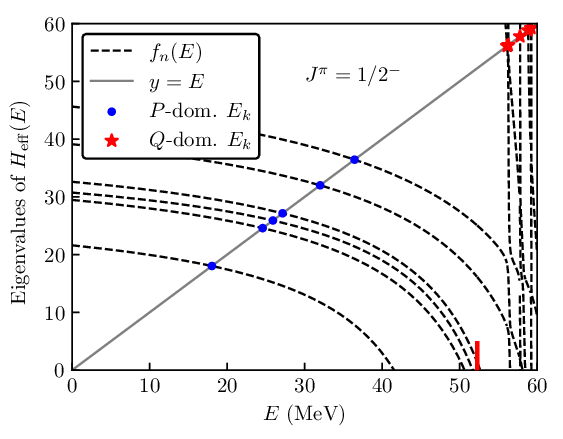} 
\caption{\label{fig:Li7_N3LObare_hw18_Nm2_HO_Exact}Same as Fig.~\ref{fig:Li7_DJ16_hw18_Nm2_HO_Exact} for $J^\pi=1/2^-$ of $^7$Li but with the bare N$^3$LO potential.}
\end{figure}
\begin{figure}[h]
\centering
\includegraphics[width=0.45\textwidth]{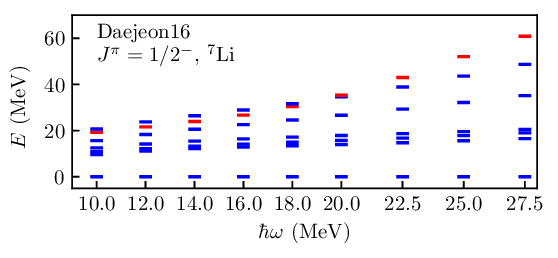} 
\includegraphics[width=0.45\textwidth]{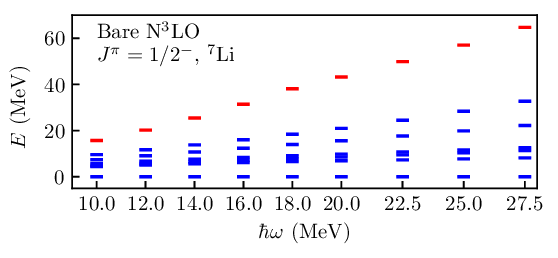} 
\caption{\label{fig:Li7_DJ16_N3LObare_NCSM_Nm2}NCSM calculations for seven lowest $J^\pi=1/2^-$ states of $^7$Li using the Daejeon16 potential and the bare N$^3$LO potential, at $N_{\rm max}=2$. The lowest $\mathbbm{Q}$-space dominated state is marked in red.}
\end{figure}

\section{Results}\label{sec:results}

In this section we apply the above BW formulation of MBPT to open-shell nuclei $^{5,6,7}$Li to illustrate the convergence criterion at work. The high order terms of the BW perturbation series of the effective Hamiltonian are calculated from the $\hat{K}$-box iterative method. We use the exact matrix inversion method and the NCSM to benchmark the results from perturbative calculations. Two nucleon-nucleon potentials, Daejeon16~\cite{DJ16} and bare N$^3$LO~\cite{EnMa2003}, with additional Coulomb potential for protons, are used in the numerical calculation. The Daejeon16 potential is based on a similarity renormalization group (SRG)~\cite{Bogner2010SRG} evolved N$^3$LO potential~\cite{EnMa2002,EnMa2003} and hence is a soft potential. The $M$-scheme HO many-body basis and $N_{\rm max}$-truncation are used, the same as the basis used in NCSM. We take the $0\hbar\omega$ model space as the $\mathbbm{P}$-space. In this way, the center-of-mass motion is in the ground-state for $\mathbbm{P}$-space basis, and therefore the eigenvalues of $H_{\rm eff}(E)$ are also characterized by zero CM excitation quanta. This is due to the fact that the intrinsic Hamiltonian $H$ commutes with the center-of-mass Hamiltonian, and the presence of $PHQ$ and $QHP$ operators in Eq.~(\ref{eq:p_space_schrodinger_equation}) makes sure that nonzero contributions from the $\mathbbm{Q}$-space have the same CM motion quantum numbers as the $\mathbbm{P}$-space. 

\subsection{$^5$Li} 

Figure~\ref{fig:Li5_DJ16_hw18_Nm2_HO_matrix_inversion}(a) depicts the eigenvalue functions $f_0(E)$ ($J^\pi = 1/2^-, 3/2^-$) of $H_{\rm eff}(E)$ for $^5$Li calculated from the exact matrix inversion method using the Daejeon16 potential at $\hbar\omega=18$~MeV, $N_{\rm max}=2$. For $^5$Li, the $\mathbbm{P}$-space is two-dimensional characterized with $J^\pi = 1/2^-, 3/2^-$. For each $J^\pi$, the $\mathbbm{P}$-space is one-dimensional. We see that the first derivative of the eigenvalue $f_0(E)$ is negative, as discussed in Sec.~\ref{sec:p_space_eigenvalue_problem}. The intersections between $y=f_n(E)$ and $y=E$, marked with blue dots and red stars in the figure, are the solutions of the $\mathbbm{P}$-space eigenvalue problem Eq.~(\ref{eq:p_space_schrodinger_equation}), which should reproduce the results from NCSM calculations. Indeed, we find that these intersections can be exactly reproduced with NCSM calculations. The NCSM calculation shows that the states of blue dots are $\mathbbm{P}$-space dominated and red stars $\mathbbm{Q}$-space dominated, which coincides with the property of the eigenvalues $f_n(E)$ addressed in Sec.~\ref{sec:p_space_eigenvalue_problem}, i.e., $|f_n'(E_k)|<1$ and $|f_n'(E_k)|>1$ correspond to $\mathbbm{P}$- and $\mathbbm{Q}$-space dominated states respectively. Indeed, we observe at blue dots the slopes of $f_0(E)$ are small and at red stars the slopes are large. Another feather of $f_0(E)$ is that it exhibits singularities, which are nothing but the eigenvalues of the $QHQ$ operator in the $\mathbbm{Q}$-space. The lowest eigenvalue $E_{\rm min}^{qhq}$ of $QHQ$ for each $J^\pi$ is marked with the bottom red vertical line. Recall that for energy $E<E_{\rm min}^{qhq}$, we can always make the BW perturbative calculation of $H_{\rm eff}(E)$ converging by choosing large enough diagonal entries of $\xi$. As shown in the figure, the two lowest $J^\pi$ states are both blow the corresponding first singularities, which is guaranteed by the variational principle. This means that the BW perturbative calculation of these two states can get converged. We will see this later. A similar calculation with the bare N$^3$LO potential is plotted in Fig.~\ref{fig:Li5_DJ16_hw18_Nm2_HO_matrix_inversion}(b). The properties of $f_0(E)$ are similar to the case of Daejeon16. Since the bare N$^3$LO potential is not softened, the eigenenergies of $^5$Li calculated with the N$^3$LO potential are higher that those calculated with the Daejeon16 potential. Nevertheless, for each $J^\pi$ the lowest state is still lower than the first singularity. As discussed before, this comes from the variational principle, and is independent of the softness of the potential. Therefore, we expect converging BW perturbative calculations for these two lowest $J^\pi$ states. 

Now let us go to the BW perturbative calculation. As we expected, the BW perturbation series for the effective Hamiltonian $H_{\rm eff}(E)$ is converging in the energy interval $E<E_{\rm min}^{qhq}$ with large enough diagonal entries of $\xi$. Equation~(\ref{eq:optimized_xi}) suggests that all the diagonal entries of $\xi$ (i.e., $\xi_k^q$) in the $\mathbbm{Q}$-space should be larger than $-144.782$~MeV and $-149.936$~MeV for $J^\pi=1/2^-$ and $J^\pi=3/2^-$ respectively. We here simply choose $\xi_k = \langle{\rm C}|H_1|{\rm C}\rangle=-130.227$~MeV, i.e., $\xi$ is a scalar, where $|{\rm C}\rangle$ is the Slater determinant with the lowest energy of the core nucleus $^4$He. This choice of $\xi$ is similar to the M{\o}ller-Plesset partitioning~\cite{MP} with a normal-ordered Hamiltonian, commonly used in the RS formalism of MBPT~\cite{Tichai2020}. In Fig.~\ref{fig:Li5_DJ16_hw18_Nm2_HO_Part1_Kbox_Nmax2}, we show the eigenvalues $f_0^{s{\rm th}}(E)$ ($J^\pi = 1/2^-, 3/2^-$) of $H_{\rm eff}^{s{\rm th}}(E)$ constructed from the BW perturbation series up to various orders $s$ for $^5$Li using the Daejeon16 potential at $\hbar\omega=18$~MeV, $N_{\rm max}=2$. The spectral radius $\rho(R)$ is also plotted for each $J^\pi$. We observe that in the energy intervals $E<10.152$~MeV for $J^\pi=1/2^-$ and $E<4.299$~MeV for $J^\pi=3/2^-$, the convergence criterion $\rho(R)<1$ are satisfied. The energies $10.152$~MeV and $4.299$~MeV are nothing but the lowest eigenvalues of $QHQ$ operator with $J^\pi=1/2^-$ and $J^\pi=3/2^-$ respectively, i.e., the lowest singularities of $f_0(E)$. Therefore the perturbative calculations inside the convergence intervals should get converged. Indeed, we observe for each $J^\pi$ the eigenvalue $f_0^{s{\rm th}}(E)$ is converging to the exact $f_0(E)$ calculated from the matrix inversion method in the interval $E<E_{\rm min}^{qhq}$. The lowest states (intersections) of $J^\pi=1/2^-$ and $J^\pi=3/2^-$ are both in the convergence interval, and thus can be obtained via converging BW perturbative calculations. The intersections residing outside the convergence interval cannot be obtained via BW perturbative calculations. The calculation with the bare N$^3$LO potential is similar to the case of Daejeon16. 

The above calculations are done at $N_{\rm max}=2$ for illustrative purpose. In practice, we use the Newton-Raphson method to quickly locate the position of the intersections between $y=f_n(E)$ and $y=E$. Figure~\ref{fig:Li5_DJ16_hw18_Nm2_HO_Kbox_Nmax246}(a) shows the BW perturbative calculation of the two lowest states ($J^\pi=1/2^-,3/2^-$) of $^5$Li using the Daejeon16 potential with $\xi=\langle{\rm C}|H_1|{\rm C}\rangle=-130.227$~MeV, at $\hbar\omega=18$~MeV, $N_{\rm max}=2, 4, 6$. The NCSM calculation with the same potential and model space is also performed to provide benchmarks. As discussed previously, the BW perturbative calculation of these two lowest $J^\pi$ states is converging with large enough $\xi$ due to the variational principle. We see that the choice of $\xi=\langle{\rm C}|H_1|{\rm C}\rangle=-130.227$~MeV is enough to make the calculations converging to the NCSM results. Varying the value of $\xi$ is able to tune the convergence speed of the BW perturbative calculation. Here we notice that as $N_{\rm max}$ increases, the speed of convergence becomes lower and lower. This may be explained as follows. The size of the $\mathbbm{Q}$-space increases as we increase the size of the full model space (characterized by $N_{\rm max}$), since we choose the $0\hbar\omega$ model space as the $\mathbbm{P}$-space. Larger $\mathbbm{Q}$-space gives lower eigenvalues of $QHQ$, and hence pushes the lowest eigenvalue characterized by $J^\pi$ of $QHQ$ closer to the lowest $J^\pi$ eigenvalue of the Hamiltonian in the full model space (i.e., pushes the lowest singularity of $f_n(E)$ closer to the lowest intersection between $y=E$ and $y=f_n(E)$ for a given $J^\pi$). Therefore the value of $\rho(R)$ at the lowest intersection gets larger but cannot exceed one, which slows down the speed of convergence. If we continue to increase $N_{\rm max}$, $\rho(R)<1$ still holds at the lowest intersection. This is assured by the variational principle and is reflected by the feature of negative slope of $f_n(E)$. The calculations with the bare N$^3$LO potential and $\xi=\langle{\rm C}|H_1|{\rm C}\rangle=-103.136$~MeV are plotted in Fig.~\ref{fig:Li5_DJ16_hw18_Nm2_HO_Kbox_Nmax246}(b). As we expected, even with the unsoftened N$^3$LO potential, the BW perturbative calculation of each lowest $J^\pi$ state can still get converged with a proper $\xi$. Furthermore, the speed of convergence is even faster than the case of Daejeon16, with the same choice of $\xi=\langle{\rm C}|H_1|{\rm C}\rangle$. This is more obvious in the calculations of $^6$Li and $^7$Li, which will be given later. 

\subsection{$^6$Li} 

The case of $^6$Li in the $p$-shell provides us with the first example of $\mathbbm{P}$-space with $d_p>1$ for a given $J^\pi$. The eigenvalue functions $f_n(E)$ of $H_{\rm eff}(E)$ for $J^\pi=0^+, 1^+, 2^+, 3^+$ calculated from the exact matrix inversion method using the Daejeon16 potential at $\hbar\omega=18$~MeV, $N_{\rm max}=2$ are plotted in Fig.~\ref{fig:Li6_DJ16_hw18_Nm2_HO_matrix_inversion}. The properties of these eigenvalues $f_n(E)$ are similar to the case of $^5$Li discussed above, except that now we have multi-dimensional $\mathbbm{P}$-space for $J^\pi=0^+, 1^+, 2^+$. We observe eigenvalue ``crossings'' for multi-dimensional $\mathbbm{P}$-space in this figure, as marked with the gray rectangle shadow, for example. These ``crossings'' are actually avoided level crossings, as shown in Fig.~\ref{fig:Li5_DJ16_hw18_Nm2_HO_Exact_crossing_details} for the marked ``crossing'', for example. The intersections between the eigenvalue functions $y=f_n(E)$ and $y=E$ are the solutions of the $\mathbbm{P}$-space eigenvalue problem Eq.~(\ref{eq:p_space_schrodinger_equation}). These intersections can exactly reproduce the corresponding NCSM calculations. The energies of $\mathbbm{P}$- and $\mathbbm{Q}$-space dominated states are marked with blue dots and red stars respectively. Correspondingly, the slopes of $f_n(E)$ are small at blue dots and are large at red stars. From the discussions in Sec.~\ref{sec:convergence_criterion}, we expect that the perturbative calculation for these marked states below the first singularities can get converged with large enough $\xi$. As an example, Fig.~\ref{fig:Li6_DJ16_hw18_Nm2_HO_Part1_Kbox_Nmax2} shows the eigenvalue functions $f_n^{s{\rm th}}(E)$ of $H_{\rm eff}^{s{\rm th}}(E)$ for $J^\pi=0^+$ from the BW perturbative calculation up to various orders $s$. The corresponding values of $\rho(R)$ at various energies are plotted in the top panel. We see that the perturbative calculations in the convergence interval $E<E_{\rm min}^{qhq}=4.621$~MeV (where $\rho(R)<1$) can indeed converge to the results from the exact matrix inversion calculation. There are two intersections residing in the convergence interval, which correspond to the two lowest $0^+$ states and the BW perturbative calculation for these two states can get converged. For the intersections (states) residing outside the convergence interval, the BW perturbative calculation cannot get converged. 

The BW perturbative calculations using the Daejeon16 potential for the states ($J^\pi=0^+, 1^+, 2^+, 3^+$) residing inside the convergence intervals at $N_{\rm max}=2, 4, 6$ are plotted in Fig.~\ref{fig:Li_DJ16_hw18_Nm2_HO_Kbox}. $\xi = \langle{\rm C}|H_1|{\rm C}\rangle=-128.427$~MeV is used in the calculation. The corresponding NCSM calculations are also performed to provide benchmarks. We see that the convergence behavior of the perturbative calculations is similar to the case of $^5$Li. We have now more than one state for $J^\pi=0^+, 1^+, 2^+$. The BW perturbative calculation for the lowest state of each $J^\pi$ is converging due to the variational principle. Although the perturbative calculations of the excited states for each $J^\pi$ are all converging to the NCSM results in this figure, the convergence of the excited state calculation cannot be guaranteed for higher $N_{\rm max}$ or other potentials, since an intruder state might occur and spoil the converging series for these excited states. Figure~\ref{fig:Li6_N3LObare_hw18_Nm2_HO_Kbox} plots the similar calculations for $^6$Li with the bare (unsoftened) N$^3$LO potential and $\xi = \langle{\rm C}|H_1|{\rm C}\rangle=-101.336$~MeV. We observe that all the calculations with the bare N$^3$LO converge to the NCSM results, and the speed of convergence is even faster than in the case of calculations with the Daejeon16 potential with the same choice of $\xi = \langle{\rm C}|H_1|{\rm C}\rangle$. We will also see this feature in the case of $^7$Li in the next subsection.  

\subsection{$^7$Li} 

Figure~\ref{fig:Li7_DJ16_hw18_Nm2_HO_Exact} presents the eigenvalue functions $f_n(E)$ ($J^\pi = 1/2^-, 3/2^-, 5/2^-, 7/2^-$) of $H_{\rm eff}(E)$ for $^7$Li calculated by the exact matrix inversion method using the Daejeon16 potential at $\hbar\omega=18$~MeV, $N_{\rm max}=2$. The result is similar to the case of $^6$Li. However, for $J^\pi=1/2^-$ only 5 intersections between $y=f_n(E)$ and $y=E$ are below the lowest singularity, whereas the $\mathbbm{P}$-space is 6-dimensional. We notice that there is an intruder state ($\mathbbm{Q}$-space dominated) below the highest $\mathbbm{P}$-space dominated state. The presence of this intruder state (or the slightly lower singularity) hinders the convergence in the BW perturbative calculation, which will be seen later. The corresponding BW perturbative calculation for $J^\pi=1/2^-$ is shown in Fig.~\ref{fig:Li7_DJ16_hw18_Nm2_HO_Kbox_Nmax246} with $\xi = \langle{\rm C}|H_1|{\rm C}\rangle=-127.141$~MeV. Indeed, we observe that the BW perturbative calculations are converging to the NCSM results expect for the highest $J^\pi=1/2^-$ state, at $N_{\rm max}=2$. The difference between the result of MBPT and NCSM changes the sign at 60th order for this highest $J^\pi=1/2^-$ state at $N_{\rm max}=2$. At $N_{\rm max}=4$, this change of sign will be present at higher orders because of the slower speed of convergence. We also observe that for a given $J^\pi$, as the energy of the state increases, the convergence speed slows down until the divergence is present, which corresponds to the fact that the value of $\rho(R)$ becomes larger and larger and eventually exceeds 1. However there is an anomaly for the $J_k^\pi=1/2_3^-$ state, whose speed of convergence is faster than that of $J_k^\pi=1/2_2^-$. This is due to the approximate isospin symmetry, namely $T\simeq3/2$ for the $J_k^\pi=1/2_3^-$ state, whereas $T\simeq1/2$ for the other $J^\pi=1/2^-$ states. The same perturbative calculation with the bare N$^3$LO potential is shown in Fig.~\ref{fig:Li7_N3LObare_hw18_Nm2_HO_Kbox_Nmax246}. Surprisingly, the perturbative calculations of all six $J^\pi=1/2^-$ states are quickly converging to the NCSM result, compared to the case of Daejeon16. The corresponding exact eigenvalue functions $f_n(E)$ ($J^\pi=1/2^-$) of $H_{\rm eff}(E)$ as a function of $E$ is depicted in Fig.~\ref{fig:Li7_N3LObare_hw18_Nm2_HO_Exact}. From this figure, we see that the gap between the $\mathbbm{P}$-space dominated states and the $\mathbbm{Q}$-space dominated states (and also the lowest singularity) is pretty large and no intruder states are present, in contrast to the case of Daejeon16. This may be explained as follows. The SRG evolution makes the NCSM calculation converging faster with respect to the size of the model space, and therefore the intruder state can occur at lower energy even in a small model space. To illustrate this idea, we plot in Fig.~\ref{fig:Li7_DJ16_N3LObare_NCSM_Nm2} seven lowest $J^\pi=1/2^-$ states of $^7$Li, calculated with NCSM at $N_{\rm max}=2$, using the Daejeon16 and bare N$^3$LO potentials (upper and lower panels, respectively). As seen from the figure, the lowest $\mathbbm{Q}$-space dominated state becomes an intruder state at $\hbar\omega<20$~MeV using the Daejeon16 potential (it appears at lower energies than the highest $\mathbbm{P}$-space dominated state). At the same time a large gap between $\mathbbm{P}$-space dominated states and the lowest $\mathbbm{Q}$-space dominated state is observed using the bare N$^3$LO potential. The presence of intruder states at low energy slows down the speed of, or even prevents, the order-by-order convergence of the BW perturbative calculation.

\section{Conclusions and Perspectives}

In summary, we generalize the novel developments of BW MBPT for the ground states of closed-shell nuclei given in Ref.~\cite{BWMBPT2023} to ground and excited states of open-shell nuclei. The general Hamiltonian partitioning scheme, the convergence criterion, and the $\hat{K}$-box iterative method for the BW perturbative calculation are introduced in the calculation of ground and excited states of open-shell nuclei. Analytical derivations show that with a large enough partitioning parameter $\xi$ (namely large enough diagonal entries of $\xi$) the BW perturbation series can always be made convergent for the states lower than the lowest eigenvalue of the Hamiltonian matrix in the excluded $\mathbbm{Q}$-space for a given $J^\pi$. It follows that it is always converging for each lowest $J^\pi$ state with large enough $\xi$, due to the variational principle. This conclusion is independent of the choice of the basis or the choice of the internucleon interaction. 

To numerically check the above conclusions, we perform proof-of-principle calculations for the open-shell nuclei $^{5,6,7}$Li using the (soft) Daejeon16 potential and the (hard) bare N$^3$LO potential in the HO basis, benchmarked with the results from the exact matrix inversion method and NCSM. We simply choose $\xi = \langle{\rm C}|H_1|{\rm C}\rangle$ in these calculations. From these calculations we can conclude: 
\begin{enumerate}
\item[(a)] The BW perturbative calculation for each lowest $J^\pi$ state is always converging with both soft and hard potentials in the HO basis with $\xi = \langle{\rm C}|H_1|{\rm C}\rangle$. This conclusion is different from the RS perturbative calculations given in Refs.~\cite{Roth2010,Langhammer2012_Pade}, where the perturbation series is diverging in the HO basis even with soft potentials. 
\item[(b)] The speed of convergence of BW perturbative calculation becomes lower as the size of full model space increases, and for excited $J^\pi$ states the convergent BW perturbation series can become divergent due to the presence of intruder states, especially in large model space.
\item[(c)] The BW perturbative calculation with hard (bare N$^3$LO) potential converges faster than with soft (Daejeon16) potential using the same choice of $\xi = \langle{\rm C}|H_1|{\rm C}\rangle$ at same $N_{\rm max}$. 
\end{enumerate}

With the developments in this paper, many works can be done in the future. For example, with the iterative $\hat{K}$-box method, the energy levels and other physical observables for other open-shell nuclei can also be calculated. However, for open-shell nuclei not very close to the closed-shell, the $\mathbbm{P}$-space dimension is large, which makes this kind of calculations less efficient. To overcome this problem, we can use the calculations for one- and two-valence nucleon systems to construct effective Hamiltonians used in valence-space shell model calculations, so that the open-shell nuclei can be calculated efficiently. Another problem is that the use of the $M$-scheme many-body basis in the $\hat{K}$-box iterative calculation may prohibit reaching heavier nuclei, similar to the case of NCSM. Instead of this algebraic iterative $\hat{K}$-box method, we can formulate the $\hat{K}$-box iterative equation~(\ref{eq:kbox_iterative_equation}) in a (Feynman) diagrammatic form so that particle-hole excitations can be symmetrically truncated. This diagrammatic method will be a brand new nonperturbative \textit{ab initio} many-body method with the ability to reach intermediate-mass and even heavy nuclei. The three-body forces can be introduced in the algebraic iterative $\hat{K}$-box method in the way as NCSM does. In the diagrammatic $\hat{K}$-box calculation, the normal-ordered approximation up to two-body level for the three-body forces can be used. Besides, the partitioning parameter $\xi$ can also be introduced to the RS perturbation theory, especially in the diagrammatic formalism, to tune the convergence behavior. These developments are under way and the results will be published separately.

\acknowledgments{%
The authors acknowledge the financial support from CNRS/IN2P3, France, via ENFIA and ABI-CONFI Master projects. 
Large-scale computations have been performed at MCIA, University of Bordeaux.
} 

\bibliography{main_bwpt_open_shell}

\end{document}